# Spin-orbit torque switching of magnetic tunnel junctions for memory applications

Viola Krizakova[1,*], Manu Perumkunnil[2], Sébastien Couet[2], Pietro Gambardella[1,*], and Kevin Garello[3,*]

[1] *Department of Materials, ETH Zurich, 8093 Zurich, Switzerland*
[2] *IMEC, Kapeldreef 75, B-3001 Leuven, Belgium*
[3] *Univ. Grenoble Alpes, CEA, CNRS, Grenoble INP, SPINTEC, 38000 Grenoble, France*

## Abstract

Spin-orbit torques (SOT) provide a versatile tool to manipulate the magnetization of diverse classes of materials and devices using electric currents, leading to novel spintronic memory and computing approaches. In parallel to spin transfer torques (STT), which have emerged as a leading non-volatile memory technology, SOT broaden the scope of current-induced magnetic switching to applications that run close to the clock speed of the central processing unit and unconventional computing architectures. In this paper, we review the fundamental characteristics of SOT and their use to switch magnetic tunnel junction (MTJ) devices, the elementary unit of the magnetoresistive random access memory (MRAM). In the first part, we illustrate the physical mechanisms that drive the SOT and magnetization reversal in nanoscale structures. In the second part, we focus on the SOT-MTJ cell. We discuss the anatomy of the MTJ in terms of materials and stack development, summarize the figures of merit for SOT switching, review the field-free operation of perpendicularly magnetized MTJs, and present options to combine SOT, STT and voltage-gate assisted switching. In the third part, we consider SOT-MRAMs in the perspective of circuit integration processes, introducing considerations on scaling and performance, as well as macro-design architectures. We thus bridge the fundamental description of SOT-driven magnetization dynamics with an application-oriented perspective, including device and system-level considerations, goals, and challenges.

## 1. Introduction

Spin torques arise from the interaction of the conduction electrons' spins with the local magnetic moments of a magnetic material [1–7]. Thanks to this interaction, mediated by exchange coupling, an electric current can torque the magnetization and induce, e.g., magnetic switching, domain wall motion or magnetic oscillations. For the memory technology, the electrical control of the magnetization in magnetic tunnel junctions (MTJs) represents a significant advancement, as it offers speed, nonvolatility, reliability, efficiency of control, and scalability unparalleled by other means [8–11].

The continuous growth in the demand to store, process, and access data is one of the main drivers of spintronic research. Emerging big data and AI computing technologies require writing data in less time and storing them at a smaller scale, and consuming less energy. Over time, computer memories based on CMOS transistors evolved into a many-level hierarchy, in which the operation speed and storage density, hence also cost, go against each other [see Fig. 1(a)]. As the physical limits of CMOS memories are being approached, the efforts to find alternatives have intensified [12,13]. A particularly interesting candidate is the magnetoresistive random access memory (MRAM), the base unit of which is the MTJ [8,14–16]. The core of an MTJ is a pair of nanomagnets separated by a thin insulating barrier. One of these magnets is pinned and serves as a "reference", whereas the other is "free" to change direction upon excitation. In this way, the MTJ can permanently store the memory bit in the relative alignment of the magnetic moments of the two magnets: parallel (P) or antiparallel (AP). Thanks to the tunnel magnetoresistance effect (TMR) [17–20], the electrical resistance of the MTJ depends on the layers' magnetic alignment, and hence the bit information can be read electrically with ease. Writing can be achieved by various means including the use of a magnetic field, electric field, and current-induced torques, which differ in efficiency, speed, and endurance, and set the requirements for the design of the memory cell [Fig.1(b)]. Overall, the MTJs are versatile devices, which can complement the embedded memory hierarchy at different levels [see Fig. 1(a)] and bridge the gap between computing and memory – a concept known as in-memory computing [21,22].

Owing to its high TMR [23,24] and interfacial perpendicular magnetic anisotropy (PMA) – qualities important for good readability and size reduction [1,25] – the CoFeB/MgO/CoFeB system has become the basis of most MTJ designs. Spin transfer torque (STT) [3–5] has meanwhile become the dominant writing mechanism thanks to its local action and scalability. The STT cell benefits from competitive power requirements and a compact design [Fig. 1(b)], in which the reading and writing paths are shared [9,16]. Notably, high-performance STT-MTJ cells are included in the first MRAMs that have recently been commercialized as embedded flash memory and last-level cache replacement. However, the shared reading/writing path also increases susceptibility to erroneous writing, read disturbance, and barrier degradation. These effects become prominent at high operation frequencies, which effectively limits the writing latency to > 5 ns. Recently, spin-orbit torques (SOT) have attracted interest as a possible alternative to STT suitable for high-speed and low error rate operation [2,26–31]. Unlike the STT-MTJ, the SOT-MTJ device has three terminals and separate reading and writing paths. Separating the writing current path from the tunnel barrier into an adjacent nonmagnetic layer with strong spin-orbit interaction, prolongs the device life-span even when operated at sub-nanosecond timescales. Whereas fundamental research efforts focus on improving the SOT figures of merit in novel materials and heterostructures, engineering efforts aim at optimizing the bit cell integration and circuit design in order to take advantage of the versatility and speed of SOT devices.

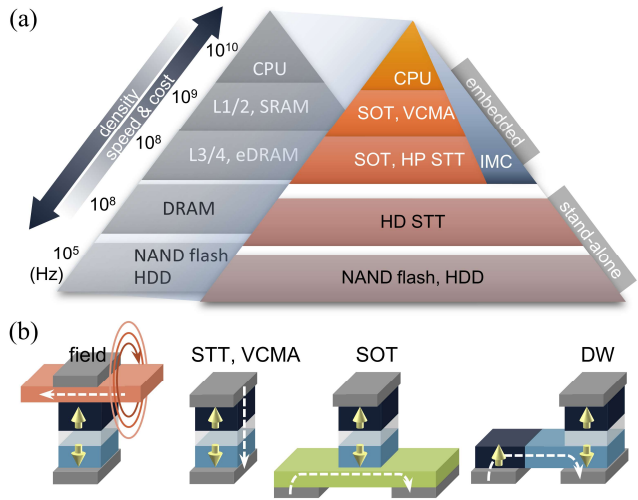

**Fig. 1.** (a) Different memory types and prospective spintronic technologies that could replace them. L1-4 stands for different cache levels, SRAM and DRAM for static and dynamic random access memory, respectively, HDD for magnetic hard disk drives. HP and HD STT are high-performance and high-density STT-MRAM, respectively, IMC stands for in-memory computing. (b) Writing mechanisms in MTJ-based devices. From left to right: Oersted field, spin transfer torque and voltage-control of magnetic anisotropy (VCMA), spin-orbit torque, and domain-wall (DW) motion.

This review provides an overview of SOT-induced magnetization reversal and its application in MRAM devices, including integration and system-level developments as well as future opportunities for in-memory computing and neuromorphic applications. The text is organized as follows: The basic principles of spin torque generation are described in Section 2 and the SOT switching dynamics is presented in Section 3. Section 4 reviews the SOT-MTJ geometries with an emphasis on the zero-external-field operation and the progress in material optimization. Section 5 summarizes the experimental realization and figures of merit for SOT switching and device options for SOT switching assisted by STT and voltage control of magnetic anisotropy (VCMA). Section 6 reviews the integration processes of SOT-MTJs, system-level considerations, and related challenges. Section 7 provides an outlook for beyond-MRAM applications of SOT-MTJs.

## 2. Fundamentals of SOT

### 2.1 Charge-spin conversion and generation of SOT

SOT are a manifestation of the relativistic spin-orbit interaction (SOI) that couples the orbital electron motion to the electron's spin [32]. On the microscopic level, their origin can be explained by different mechanisms [2,6]. When a charge current is passed through a nonmagnetic conductor, the SOI mediates its conversion into a transverse spin current that propagates toward the interfaces and where it results in spin accumulation. This mechanism is known as the spin Hall effect and is prominent in heavy metals, such as Pt, Ta, and W [33,34]. In the presence of broken inversion symmetry, e.g., at the conductor's interface, electrons additionally experience an electric field perpendicular to the interface plane, which acts as a magnetic field in the electron's rest frame and results in non-equilibrium spin accumulation at the interface. This mechanism is known as the Rashba-Edelstein effect [7,35–37], and is prominent at the interface of a heavy metal with another metal having different electronegativity [38,39], as well as at oxide interfaces [40–42], and low-dimensional semiconductors [43]. The Edelstein effect is also responsible for charge-spin conversion in materials with spin-momentum locked electron states, such as topological insulators and Weyl semimetals [37,44,45]. A combination of these effects is also possible, in conjunction with additional effects due to electron scattering at the interface between magnetic and nonmagnetic conductors and the generation of spin currents within a magnetic material [46–48]. Recently, orbital analogues of the spin Hall effect and Rashba-Edelstein effects [49–52] have been proposed as additional mechanisms for the generation of SOT, provided that the SOI converts an orbital current into a spin current inside or outside the magnetic layer [53–56]. When a magnetic material is deposited on top or below a nonmagnetic layer with SOI-induced charge-spin conversion, the spin current produced by any of the mechanisms mentioned above will diffuse into the magnet and torque the magnetization, as schematized in Fig. 2(a). Overall, the multiplicity of effects that are able to generate SOT allows for a wide choice of materials and optimization strategies to maximize their strength and efficiency for magnetization switching [2].

In general, a spin-orbit torque $\boldsymbol{\Gamma}$ can be decomposed into two perpendicular components [Fig. 2(b)]: a longitudinal dampinglike torque (DLT) acting on the magnetization as an effective damping term, and a transversal fieldlike torque (FLT) acting on the magnetization as an effective magnetic field of fixed orientation. Together, they can be expressed as

$$\boldsymbol{\Gamma} = \tau_{\text{DL}} \boldsymbol{m} \times (\boldsymbol{m} \times \boldsymbol{\sigma}) + \tau_{\text{FL}} (\boldsymbol{m} \times \boldsymbol{\sigma}) \quad (1)$$

where $\boldsymbol{m}$ is the unit magnetization vector, $\boldsymbol{\sigma}$ is the polarization of the spin current, and $\tau_{\text{DL,FL}}$ is the amplitude of the respective torque component [2,57,58]. In most cases, $\boldsymbol{\sigma}$ is oriented in-plane perpendicular to the direction of both the charge and spin current, although low-symmetry materials allow also for out-of-plane $\boldsymbol{\sigma}$ components [59–61].

For SOT switching, the DLT is more relevant, as it destabilizes the magnetization and induces its rotation toward $\boldsymbol{\sigma}$, which can lead to the reversal, whereas the FLT drives the magnetization into precession and impacts the magnetization dynamics and the reversal speed. In the spin Hall effect picture, the amount of DLT generated by the charge current $j$ and acting on a ferromagnet with thickness $t_{\text{FM}}$ can be expressed as $\tau_{\text{DL}} = \frac{\theta_{\text{SH}}}{M_s t_{\text{FM}}} \frac{\hbar}{2e} j$, where the effective spin Hall angle $\theta_{\text{SH}}$ represents the amount of spin current absorbed by the ferromagnet per unit of charge current. This parameter is often used as a proxy for the proper spin Hall angle of the nonmagnetic conductor, namely the ratio of the spin Hall-to-electrical conductivity. However, spin back-flow, as well as the spin transparency of the interface and spin memory loss, effectively decrease the spin current reaching the ferromagnet [62–64], so that the two parameters can differ significantly from each other. Moreover, SOT generation mechanisms other than the spin Hall effect are included in the definition of $\theta_{\text{SH}}$ because torque measurements cannot easily discriminate between different charge-spin conversion processes.

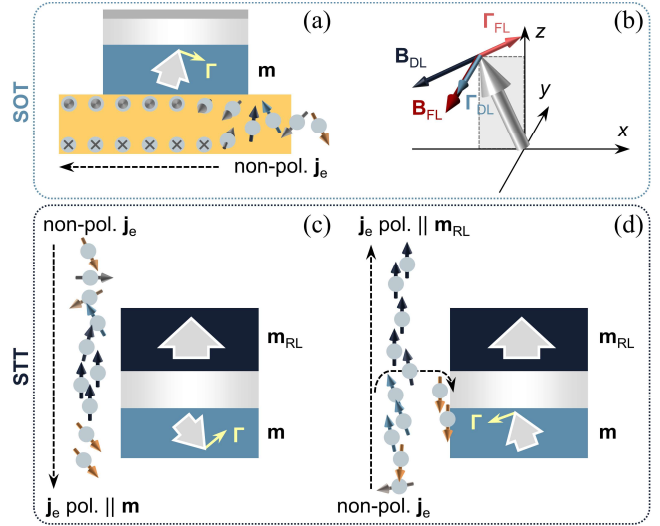

**Fig. 2**. Switching of a PMA system by spin-orbit and spin transfer torques. (a) Generation of SOT and the switching geometry in a PMA system with SOT generated by the spin Hall effect. (b) Dampinglike and fieldlike SOTs and respective effective fields induced by current $j$ along $x$. (c) Switching of the PMA MTJ by STT to the P state and (d) to the AP state. White arrows show the magnetization direction of the reference ($\boldsymbol{m}_{\text{RL}}$) and free ($\boldsymbol{m}$) magnetic layer. Black dashed arrows in (a), (c) and (d) indicate the direction of the electron flow $j_e$.

Since $\boldsymbol{\Gamma} = \boldsymbol{M} \times \boldsymbol{B}$, the DLT and FLT can also be expressed in terms of the effective magnetic fields $\boldsymbol{B}_{\text{DL}} = B_{\text{DL}}(\boldsymbol{m} \times \boldsymbol{\sigma})$ and $\boldsymbol{B}_{\text{FL}} = B_{\text{FL}} \boldsymbol{\sigma}$, which allows for comparing the effect of the current on the magnetization with that of externally applied magnetic fields. Using the effective field definition, the dimensionless SOT efficiencies are written as [2,63]

$$\xi_{\text{DL,FL}} = \frac{2e}{\hbar} \frac{M_s t_{FM}}{j} B_{\text{DL,FL}} \quad (2)$$

These parameters quantify the effective SOT magnetic fields per unit of electric current in a given system, and are therefore crucial in determining the energy consumption required for magnetization reversal. SOT efficiencies normalized by the electric field, namely $\xi^E_{\text{DL,FL}} = \xi_{\text{DL,FL}}/\rho$, have also been introduced to compare different material systems, which are equivalent to an effective spin Hall conductivity. Caution should be exerted when comparing results across different systems, however, because the current density $j$ is often, though not always, determined by considering only the current flowing in the nonmagnetic layer using a parallel resistor model, which is a rather crude approximation in very thin layers. Moreover, current shunting through the magnetic layer can be significant [2] and total current considerations are more appropriate for benchmarking devices. There are several methods to experimentally characterize spin-orbit efficiencies $\xi_{\mathrm{DL,FL}}$, a resume of them can found in Refs. [2,65].

## 2.2 Comparison with STT

STT originate from the transfer of spin-angular momentum from one magnetic layer to another, mediated by the spin-polarized current that flows with the charge current [3–5,66]. Differently from the transverse spin current giving rise to SOT, this is a longitudinal spin current with polarization set by the magnetization of the first layer. When electrons polarized along the magnetization of a ferromagnet (the reference layer) tunnel through a potential barrier into another magnet (the free layer) with a different magnetic orientation their polarization tends to align to the magnetization of the second magnet. Conservation of angular momentum results in a torque exerted by the tunneling electrons on the magnetization of the second magnet [Fig. 2(c)]. Moreover, due to spin-dependent scattering at the magnet's interface, some of the electrons are reflected back and exert torque on the first magnet. When the electrons flow from the reference to the free layer, the magnetization of the latter can switch and align parallel (P) to the former. When the electron flow is reversed, the electrons traveling from the free layer (FL) into the reference layer that have spin polarization opposite to that of the reference layer are strongly back-scattered, leading to the switching of the FL antiparallel (AP) to the reference layer [Fig. 2(d)]. The difference between forward- and backward-scattered currents results in asymmetric threshold currents for AP-P and P-AP switching.

Because the interaction between the spin current and local magnetization is common to both types of torques, STT have a similar functional form as SOT

$$\boldsymbol{T}^{\mathrm{STT}} = \tau_{\mathrm{DL}}^{\mathrm{STT}} \boldsymbol{m} \times (\boldsymbol{m} \times \boldsymbol{m}_{\mathrm{RL}}) + \tau_{\mathrm{FL}}^{\mathrm{STT}} (\boldsymbol{m} \times \boldsymbol{m}_{\mathrm{RL}}) \quad (3)$$

where $\boldsymbol{m}$ and $\boldsymbol{m}_{\mathrm{RL}}$ are the magnetization of the free and reference layer, respectively. The antidamping term is often referred to as Slonczewski's torque, with amplitude given by $\tau_{\mathrm{DL}}^{\mathrm{STT}} = \frac{\hbar \eta(\theta)}{2eM_s t_{\mathrm{FM}}} j$, where $\eta(\theta)$ is the spin polarization as a function of the relative angle between $\boldsymbol{m}$ and $\boldsymbol{m}_{\mathrm{RL}}$. The fieldlike component is typically small due to fast dephasing of transverse spins in metallic systems, and it is often neglected [1,66].

Overall, SOTs and STT can have similar effects on magnetization, but differ by their origin and geometry, and thus the type of dynamics they induce. The main distinctions are: i) STT requires a reference magnetic layer to spin-polarize the current, whereas SOT allows for switching of single-layer magnets. ii) The spin current flows perpendicular (parallel) to charge flow in SOT (STT). iii) The spin-polarization is usually orthogonal to both charge and spin current flow in SOT, whereas it can be arbitrarily set by $\boldsymbol{m}_{\mathrm{RL}}$ in STT. iv) In SOT, $|\xi_{\mathrm{DL}}|$ can be > 1, whereas in STT $|\eta(\theta)| \leq 1$. Moreover, when considering the switching of PMA systems, two additional considerations come into play. First, using STT a PMA magnet can be reversibly switched by injecting a bipolar current across the layers, and the switching polarity is fully determined by the direction of the spin current, i.e., $\boldsymbol{m}$ aligns parallel (antiparallel) to $\boldsymbol{m}_{\mathrm{RL}}$ for an electron current flowing from (to) the reference layer. On the contrary, using SOT an additional symmetry-breaking mechanism – usually in the form of a static in-plane magnetic field – is required to deterministically switch a layer with PMA (see Section 3) and the polarity of switching is determined by the relative alignment of the current and the magnetic field applied parallel or antiparallel to it. Second, STT is nominally zero in the quiescent state because $\boldsymbol{m}$ and $\boldsymbol{m}_{\mathrm{RL}}$ are collinear to each other in both the P and AP configurations, which requires thermally-activated fluctuations of $\boldsymbol{m}$ in order to trigger the switching process. Conversely, the SOT is maximum at the beginning of the switching process, because the DLT and FLT are orthogonal to the out-of-plane magnetization. The differences between STT and SOT make them complementary for use in different devices or for combined use in the same device, as discussed in Section 5.

## 3. SOT-driven magnetization dynamics

In this section we concentrate on the dynamics of PMA systems, as these are the most technologically relevant for SOT-MRAM applications. The SOT switching of in-plane systems and related prospects are reviewed in Section 4.

Early experiments showed that deterministic SOT switching is driven by the DLT and requires current injection above a certain threshold and the application of a symmetry-breaking magnetic field $B_x$ collinear with the current [26]. The switching is bipolar with respect to either current or in-plane field direction, as summarized in Fig. 3(a). In positive DLT systems, such as Pt/Co, for positive $B_x$, the magnetization switches down-to-up at positive current, and up-to-down at negative current (clockwise), whereas the opposite occurs upon reversing $B_x$ (anti-clockwise) [26,67]. In negative DLT systems, such as Ta/Co and W/CoFeB, the switching is anti-clockwise (clockwise) at positive (negative) $B_x$, as shown in Fig. 3(b) [27,68,69]. This behavior and the role of $B_x$ can be easily explained using both macrospin and micromagnetic arguments.

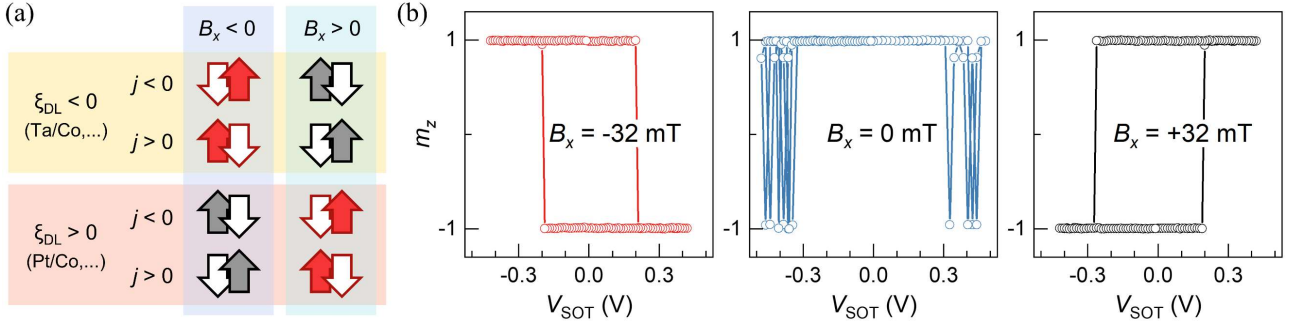

**Fig. 3.** (a) Symmetry of SOT switching in PMA systems with respect to the polarity of the field $B_x$, the current $j$, and the DLT $\xi_{DL}$. The left and right arrow represent the initial and final state of the magnetization, respectively. (b) SOT switching of a W/CoFeB sample with magnetic field $B_x$ applied against (left panel) and along (right pannel) the current $j$, and random switching without in-plane field (middle panel). (b) SOT switching of a W/CoFeB sample in the magnetic field $B_x$ applied along and against the current flow $j$, and switching without application of the in-plane field.

### 3.1 Magnetization reversal in the macrospin approximation

Macrospin simulations rely on approximating the magnetization of an extended system by a single magnetic moment and calculating its dynamics using the Landau-Lifshitz-Gilbert (LLG) equation

$$\dot{\mathbf{m}} = -\gamma \mathbf{m} \times \mathbf{B}_{eff} + \alpha \mathbf{m} \times \dot{\mathbf{m}} + \frac{\gamma}{M_s} \mathbf{\Gamma} \qquad (4)$$

where $\mathbf{B}_{eff}$ is the effective field resulting from the sum of the external magnetic field and magnetic anisotropy, $\alpha$ is the magnetic damping parameter, and $\gamma$ the electronic gyromagnetic ratio. Accordingly, the dampinglike SOT component reorients the magnetization toward the transient equilibrium direction determined by $\boldsymbol{\sigma}$ and $\mathbf{B}_{eff}$ [70,71]. If the DLT is strong compared to $\mathbf{m} \times \mathbf{B}_{eff}$, $\mathbf{m}$ tends to align antiparallel to $\boldsymbol{\sigma}$. In a PMA system, $\mathbf{m}$ is initially orthogonal to $\boldsymbol{\sigma}$ and thus reacts instantly to the excitation reaching the new equilibrium state until the current is removed, as shown in Figs. 4(a) and (b). As the current is turned off, $\mathbf{m}$ relaxes toward the out-of-plane energy minimum. If $B_x = 0$, both out-of-plane orientations are equivalent, and the switching outcome is non-deterministic. If $B_x > 0 \ (< 0)$, $\mathbf{m}$ is tilted downwards (upwards) in the transient equilibrium state, and the final state is determined by the relative alignment of $j$ and $B_x$ [26]. In this approximation, the threshold current $j_{th}$ that leads to the destabilization of the macrospin is given analytically as [71]

$$|j_{th}| = \frac{2eM_s t_{FL}}{\mu_0 \hbar \xi_{DL}} \left( \frac{B_K}{2} - \frac{B_x}{\sqrt{2}} \right), \qquad (5)$$

where the effective anisotropy field $B_K$ is assumed to be much larger than $B_x$.

The simulations shown in Figs. 4(a) and (b) neglect the FLT; however, this torque is present in a stronger or weaker form in most SOT systems [2] and has a destabilizing action on the magnetization [72]. Under the action of the FLT, the magnetization undergoes damped oscillations about the transient equilibrium direction determined by $\boldsymbol{\sigma}$ and $\mathbf{B}_{eff}$ before relaxing to it [see Figs. 4(c) and (d)]. The FLT affects the switching threshold current, which is modified as

$$|j_{th}| = \frac{2eM_s t_{FL} \sqrt{\alpha}}{\mu_0 \hbar \xi_{DL} \sqrt{\beta(2+\alpha\beta)}} \sqrt{2B_K^2 - B_x^2}, \qquad (6)$$

where $\beta$ is the ratio of FLT to DLT [73]. The macrospin model is useful to estimate the impact of the magnetic and SOT parameters on the switching threshold and outcome [73–75]; however, it significantly overestimates $j_\text{th}$ in magnetic dots and MTJs with dimensions of a few tens of nm [76–78]. The discrepancy between experimental values of the switching current and Eqs. (5,6), and its scaling with the duration of the current pulses (discussed in Section 5), indicate that SOT magnetization reversal occurs by an incoherent process that involves domain nucleation and propagation [70,79,80], as confirmed experimentally [81,82]. Future investigations of SOT switching in magnetic systems with lateral dimensions ≤ 10 nm might reveal if coherent macrospin switching is possible using SOT.

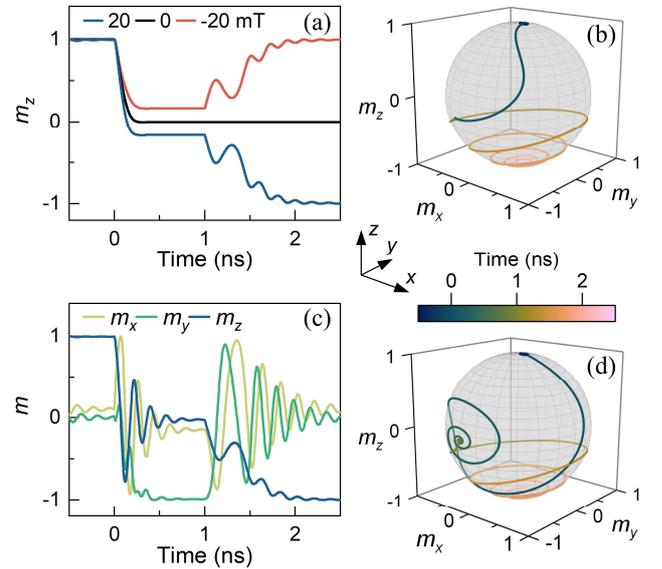

**Fig. 4.** Macrospin simulations of the time-dependent evolution of the magnetization due to SOT supplied by a 1-ns-long rectangular current pulse with $j \parallel +x$. (a) Time traces of $m_z$ for DLT, $\xi_{DL} = 0.1$, at different external fields $B_x$ and (b) the trajectory of $\mathbf{m}$ for $B_x = 20$ mT. (c) Time traces of all three components of $\mathbf{m}$ for finite DLT and FLT, $\xi_{DL}(\xi_{FL}) = 0.1 \ (0.2)$, and (d) the corresponding trajectory of $\mathbf{m}$.

## 3.2 Switching by nucleation and propagation of domain walls

Understanding incoherent SOT switching requires considering not only the effect of DLT, FLT, and $B_x$, but also the influence of the Dzyaloshinskii-Moriya interaction (DMI) and temperature on the nucleation of magnetic domains and propagation of domain walls (DW). The DMI is an antisymmetric exchange coupling that originates from SOI in systems lacking reversal symmetry, and is thus often present at the heavy metal / ferromagnet interfaces used to generate SOT [39,83–85]. Its action is equivalent to that of an effective magnetic field

$$\boldsymbol{B}_\text{DMI} = \frac{2D}{M_s}\left(-\frac{\partial m_z}{\partial x}, -\frac{\partial m_z}{\partial y}, \frac{\partial m_x}{\partial x} + \frac{\partial m_y}{\partial y}\right), \quad (7)$$

which tends to align neighboring spins orthogonally to each other following a specific (chiral) sense of rotation determined by the sign of the DMI constant $D$ [86,87]. The DMI is responsible for the formation of Néel-type DW and skyrmions in PMA systems [88,89] and for the tilting of the magnetic moments at the edge of thin magnetic structures [79,80].

The nucleation of a domain starting from a homogeneous magnetic state is triggered by the DLT and FLT, and assisted by $B_x$ and $B_\text{DMI}$, as schematized in Fig. 5(a). This process can be understood by applying the macrospin model to a reduced activation volume [90], whereby the DLT and FLT trigger the dynamics and $B_x$ and $B_\text{DMI}$ lower the energy barrier for magnetization reversal [91], as exemplified by Eq. (6). Once a reversed domain has nucleated, the DLT drives the propagation of the DW [see

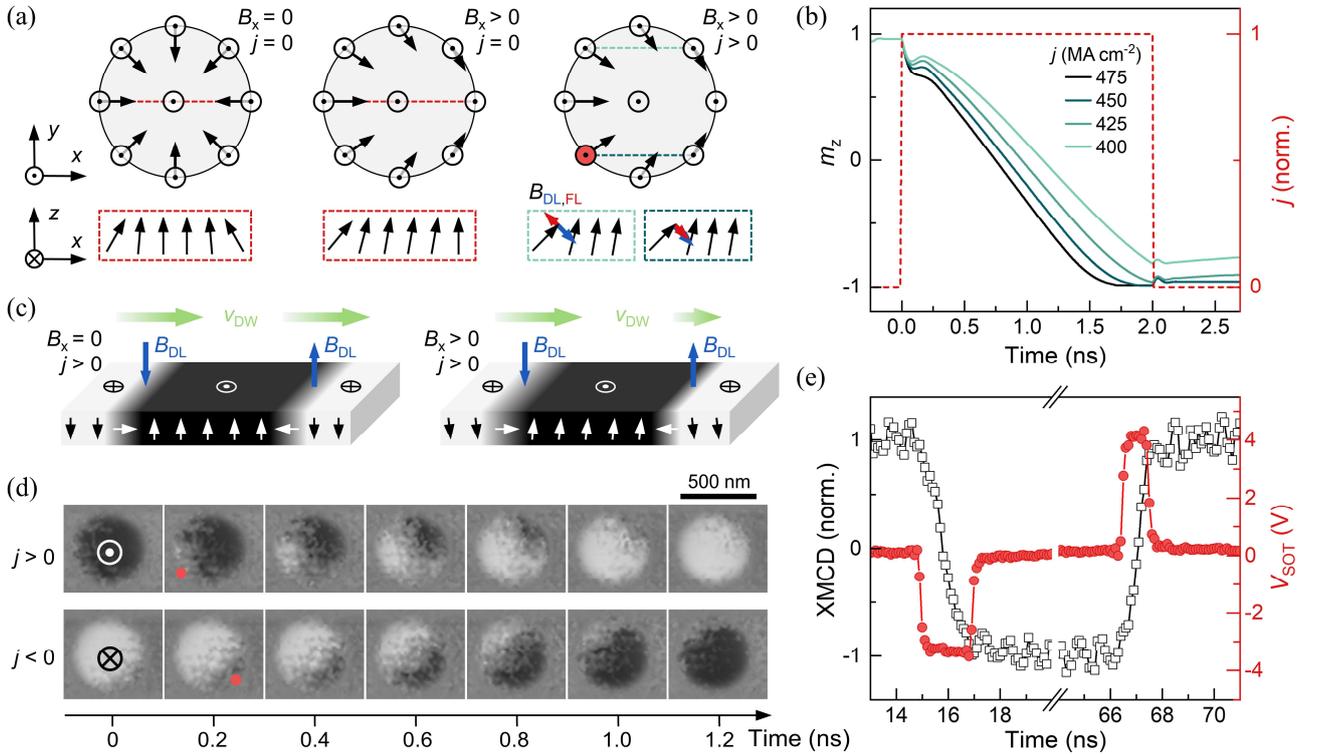

**Fig. 5.** SOT-driven magnetization reversal by domain nucleation and propagation in a system with positive $B_\text{DL}$, such as Pt/Co. (a) Schematics of the magnetization tilting around the edge of a magnetic dot due to DLT, FLT, DMI, and $B_x$ and (b) simulated evolution of the magnetization due to a 2-ns-long SOT pulse for different current amplitudes. (c) Schematics of the SOT-driven DW motion without (left) and with (right) $B_x$. (d) Magnetization frames measured by scanning transmission x-ray microscopy on a Pt/Co dot showing the nucleation site (red) and the domain expansion across the dot diagonal due to positive and negative SOT pulses with $B_x > 0$. (e) Time-resolved x-ray magnetic circular dichroism contrast corresponding to $m_z$ integrated over the dot area. Panels (a), (b), (d), and (e) are reproduced from [81].

Fig. 5(b)], as it acts on the internal magnetization of the DW as an effective easy-axis magnetic field $B_{\text{DL}} \parallel \pm z$, favoring the expansion of either an "up" or "down" domain, as shown in Fig. 5(c). For this to occur, the magnetization of the DW must have a nonzero component $m_{\text{DW},x}$ collinear with the current direction. In other words, only Néel-type or intermediate Néel-Bloch type DW can be driven by SOT, unlike Bloch-type DW [92]. The DMI is thus a crucial factor in SOT-induced DW motion, as it promotes the formation of chiral Néel DW in thin magnetic structures with PMA [93]. In this scenario, up-down and down-up DW move parallel to each other with equal velocity proportional to $B_{\text{DL}}$ [87,92]; a magnetic field $B_x$ is thus required to change $m_{\text{DW},x}$ and offset the velocity of one wall with respect to the other, leading to the net expansion of one domain, as shown in Fig. 5(c) [94,95]. Notably, contrary to STT-driven DW that move only with electron flow, SOT-driven DW move in the direction imposed by the sign of $B_{\text{DL}}$, and by the interplay of DMI and $B_x$. This opens the very interesting possibility of optimizing and manipulating DW motion not only for switching, but also for racetrack memories and domain wall logic applications [96–102].

### 3.2.1 Experimental observations of SOT switching in space and time

Time-resolved x-ray imaging with elemental and magnetic contrast confirms that SOT-switching proceeds by deterministic domain nucleation at alternating edges of a magnetic dot depending on the sign of the DLT, FLT, $B_x$ and $B_{\text{DMI}}$, and fast propagation of a DW across the dot, as shown in Figs. 5(d) and (e). This highly reproducible behavior is specific to SOT and common to ferromagnets such as Pt/Co [81] and ferrimagnets such as Pt/GdFeCo [82]. The DW are shown to tilt during motion due to simultaneous action of DLT, DMI, and, to a lesser extent, FLT [103,104]. The spatial extension of the reversed magnetic state correlates well with the amplitude of the anomalous Hall effect measured in these systems. Furthermore, switching is deterministic also in the presence of defects or multiple nucleation events [81], which makes SOT very resilient to materials and microfabrication variability issues in devices.

Time-resolved pump-probe magneto-optical Kerr effect (MOKE) measurements using ultrafast lasers allow for studying the SOT dynamics on the sub-ps timescale. MOKE studies have shown that the reversal dynamics depends on the timescale and amplitude of the SOT pulse as well as current-induced heating [105]. Whereas the DW-mediated reversal occurs upon excitation by ns-long pulses [105,106], ps-long current pulses excite nearly coherent magnetic oscillations with post-pulse dynamics lasting up to 1 ns [107,108]. Notably, a recent demonstration of reproducible SOT switching by 6 ps-long electrical pulses [107] significantly surpassed the highest switching speed observed in all-electrical devices [31], suggesting that there is extensive room for improving the SOT reversal speed in the sub-ns regime. It should be noted, however, that pump-probe detection schemes are only sensitive to reproducible dynamic features, whereas all stochastic aspects are hidden by averaging. All-electrical measurement techniques based on the TMR in MTJs [31,109] and the anomalous Hall effect in single-layer systems [110], on the other hand, provide insight into both stochastic and reproducible temporal aspects of the dynamics, albeit without spatial resolution (see Section 5).

### 3.2.2 Effects of downscaling and temperature

The structures investigated by x-rays and MOKE are typically larger than 500 nm [81,105,106] and can thus easily accommodate a DW. However, the typical SOT-MTJ pillar size is ≤ 50 nm, and usually comprises a CoFeB FL grown on $\beta$-W or Ta, in which the DMI is much weaker compared to Pt/Co bilayers [94,97]. Spatially-resolved switching in MTJs can only be studied using micromagnetic simulations. These suggest that the main characteristics of SOT switching presented above are preserved, but with some differences [31,109]. In small magnetic dots ($\lesssim 100$ nm), the DW formation is less energetically favorable than in large ones, raising the switching energy barrier and leading to more coherent magnetization reversal. Consequently, $j_{\text{th}}$ increases in smaller dots, as shown in Fig. 6(a) and confirmed experimentally [77,78], and the nucleation point of the reversed domain is less clearly defined [see Fig. 6(b)]. Moreover, a Néel wall develops a Bloch component when under the action of SOTs, which results in a reduced propagation efficiency compared to a pure Néel wall [87,92,104]. This effect is more pronounced in the weak-DMI materials.

The simulations point to other parameters that influence the switching threshold and outcome, such as the FLT, the shape of the excitation pulse, the external magnetic field, the transient temperature increase in the FL, and STT [31,75,109,111,112]. The FLT promotes oscillatory behavior, which translates to more frequent writing errors in devices, as well as a reflection of the propagating DW at the dot edge [106,113]. The FLT has been predicted also to allow for field-free switching [75] or unipolar deterministic switching [113], supposing that precise control of the FLT-to-DLT ratio can be attained. However, most 3-terminal MTJ devices use W as an SOT underlayer, which does not have a strong FLT.

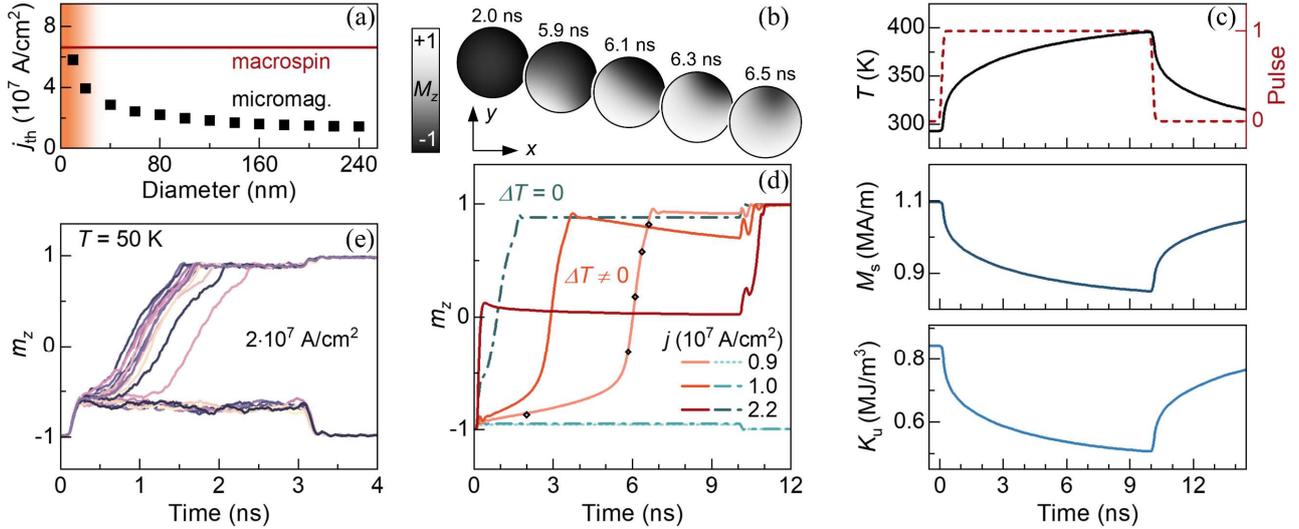

**Fig. 6.** (a) Threshold current density of SOT switching obtained from micromagnetic simulations for 2-ns-long current pulses (symbols) and macrospin limit (red line). The shaded area denotes the coherent reversal regime. (b-d) Switching of an MTJ device with parameters specified in [31] and $B_x = 20$ mT. (b) Magnetization frames corresponding to the down-to-up reversal shown in (d) at different times. (c) Temperature profile in the device during and after the SOT pulse and corresponding change of the saturation magnetization and PMA of the FL. Reproduced from [31]. (d) Evolution of the magnetization obtained from zero-temperature simulations of switching induced by three different currents: with (dashed-line traces) magnetic parameters kept constant during the simulation, or (solid-line traces) time-dependent parameters which scale as in (c). (e) Twenty finite-temperature simulations obtained including a random thermal field for the same pulse conditions and magnetic parameters kept constant during the pulse.

Temperature plays a twofold role in the switching – deterministic and stochastic. Joule heating induces a time-dependent variation of the magnetic properties of the FL, as shown in Fig. 6(c) [31], most importantly the reduction of PMA [114]. This thermal activation does not alter the reversal dynamics, but lowers the threshold current and the timing of the reversal, as illustrated in Fig. 6(d). Furthermore, stochastic thermal fluctuations, which are accounted for by including a random thermal field in micromagnetic simulations [115], result in a dispersion of the switching events, as shown in Fig. 6(e).

## 4. SOT-MRAM: from materials to MTJ devices

### 4.1 Device geometries

The unit cell of an SOT-MRAM is typically a top-pinned 3-terminal MTJ with two contacts to the SOT underlayer and one contact to the top of the pillar, which allows for electrical readout as well as controlling the STT and VCMA during SOT-writing. This design has a larger footprint compared to a 2-terminal STT cell, but separates the read and write currents, thus preserving the tunnel barrier from degradation and minimizing undesired writing errors. Notably, MTJ designs that incorporate SOT using a 2-terminal geometry [116] and multi-pillar MTJ cells sharing the same underlayer [117,118] have been proposed to increase the cell density.

The most essential requirements of the MTJ cell are related to its ability to store, read, and write the bit information while being compact and compatible with back-end-of-line (BEOL) CMOS fabrication processes. The first aspect concerns the definition of the layers' magnetic alignment and its stability against external disturbances and thermal fluctuations, commonly quantified by the thermal stability factor $\Delta$ as the ratio between the magnetic anisotropy energy of the FL and $k_B T$. Values of $\Delta > 40$ are required to ensure reliable data retention over 10 years at room temperature. The second aspect is the TMR and the breakdown voltage of the barrier, which define the power requirements and limits of the reading operation. The third aspect includes the efficiency of the charge-spin conversion as given by $\xi_{DL}$, compounded by the electrical resistance of the SOT track, which determines the power required for writing.

The different MTJ designs employing SOT writing can be categorized by the orientation of the magnetic easy axis:

i) collinear with the charge current (*x*-type) [119–121], ii) perpendicular to the current in the film plane (*y*-type) [27,30,119,122–124], and iii) out of the plane (*z*-type) [28,29,31,109,125]. The SOT switching dynamics is different in these geometries (see Table 1). The *x* and *z* device types enable very fast switching, thanks to the perpendicular alignment of the spin polarization and magnetic easy axis, which ensures the immediate onset of the torque. Moreover, the *z*-type has gained popularity for its larger thermal stability when downscaled compared to in-plane systems. A drawback common to the *x*- and *z*-type is the requirement for a symmetry-breaking mechanism to unambiguously set the final state and hence ensure deterministic bipolar operation [26]. Whereas an external magnetic field can efficiently lift the final state degeneracy, its integration remains a practical challenge (see Section 4.2).

The SOT switching of the *y*-type MTJ is similar to the STT switching. The initially collinear alignment of magnetization and the spin polarization enables deterministic switching without an additional symmetry-breaking mechanism and a steeper reduction of the threshold current with the pulse duration than its *x* and *z* counterparts, although at the expense of a significant incubation delay of the switching onset [121,123]. In the macrospin approximation, the threshold current is reduced by $\alpha$ relative to Eq. (4) as the DLT amplifies the precessional motion of ***m*** about the *y*-axis whenever ***m*** deviates away from it [71,126]. In an effort to preserve both the switching speed and field-free operation capabilities, also mixed type geometries with a tilted anisotropy axis have been proposed [121,127–130].

## 4.2 Field-free switching in perpendicularly magnetized systems

Accommodating a magnetic field source to the SOT-MRAM cell represents a problem for downscaling and device density. Substitution of a permanent magnet by integration-friendly alternatives is, therefore, a key objective in SOT switching applications, which has resulted in a broad variety of conceptually different approaches to magnetic field-free switching [2,133,134]. The proposed mechanisms rely on incorporating an effective field in the form of an embedded in-plane magnet [26,131,135], exchange bias from an antiferromagnet [136–139], interlayer coupling with a ferromagnetic layer [130,140,141], and DMI field [142]. Other approaches rely on introducing alternative symmetry breaking mechanisms, such as a wedge-shaped FL [143], tilted anisotropy axis [127], lateral asymmetry [144], strain [145], composition gradients [146], and tilted growth direction [147,148]. Finally, the generation of out-of-plane spin polarization due to STT [116,125,149], additional ferromagnetic layer [48], lateral modulation of the Rashba effect [150], or low-symmetry crystal structure [59,60,151,152] has been demonstrated, which can lead to field-free switching in combination with the conventional DLT.

These approaches differ in their strengths and limitations in relation to technological challenges such as scalability, cost of material and additional lithography steps, reliability, robustness against disturbance, compatibility with CMOS processes, as well as the preservation of large TMR and data retention. Table 2 summarizes the main characteristics of the field-free switching approaches, representative of the diverse symmetry-breaking mechanisms.

**Table 1.** Overview of the relevant switching characteristics with their target values, and comparison of the typical values achieved in the three main device geometries, *x*-, *y*-, and *z*-type, and mixed *xy* geometry.

| PARAMETER | 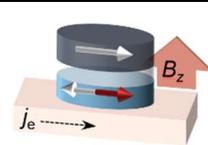 | 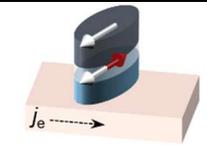 | 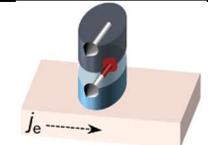 | 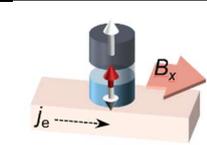 |
|---|---|---|---|---|
| Orientation | ***x*** | ***xy*** | ***y*** | ***z*** |
| Scalability | limited | limited | limited | good |
| Writing current density at 1 ns (10 ns) | 33 (30) MA/cm$^2$ | 24 (10) MA/cm$^2$ | 50 (22) MA/cm$^2$ | 126 (50) MA/cm$^2$ |
| Intrinsic critical current $j_{c0}$ | 43 MA/cm$^2$ | | 4-10 MA/cm$^2$ | 20 MA/cm$^2$ |
| $\Delta$ ($k_BT$) | 45.5 | 70 | 44-46 | 42-150 |
| Field-free operation | no | yes | yes | no |
| References | [120,121] | [128,129] | [30,120,121] | [29,120,131,132] |

**Table 2.** Overview of field-free switching demonstrations sorted by the underlying mechanisms.

| PARAMETER | Geometry | | | Built-in magnet | | z spin polarization | | | |
|---|---|---|---|---|---|---|---|---|---|
| | Lateral asymmetry | Thickness gradient | Tilted anisotropy axis | In-plane magnet | Exchange bias | STT | Rashba field modulation | Crystal symmetry | Composition gradient |
| Scalability | fair | poor | fair | good | good | good | poor | good | fair |
| Advantages | material options | material options | material options | integration | SOT design unaltered | cell density, low energy | full electrical control | low current | fast dynamics |
| Disadvantages | integration | variability, integration | fabrication, retention | additional layer, variability | growth, thermal properties | endurance | integration | growth control | integration |
| Demonstration | µ-scale device | µ-scale Hall bar | nano device | MTJ | nano device, Hall bar | MTJ | µ-scale Hall bar | µ-scale Hall bar | µ-scale Hall bar |
| References | [144] | [143,153] | [127] | [131] | [136,137,139] | [154] | [150] | [59,60,151] | [122] |

## 4.3 Fabrication and optimization of SOT-MTJ stacks

To fulfill the SOT-MRAM potential for applications, the stack design and cell layout are the first building blocks to optimize. These have to be guided by circuit-level design and system-level requirements, as discussed in Section 6, which dictate minimum specification requirements such as the writing/reading currents, retention time, and footprint. At the first level, the SOT-MTJ stack optimization relies on three interlinked aspects: i) SOT track material, focusing on its electrical conductivity to minimize the write voltage and its charge-to-spin conversion ratio $\xi_{DL}$ to minimize the write current, ii) interface between the SOT track and the FL to maximize spin current transmission, and iii) MTJ properties that will define the reading speed, the retention time and the compatibility with integration process (thermal budget, etching, contamination,…). We discuss in the following some of the aspects that are crucial to adapting SOT-MTJ performances to application needs, and progress made in the past years.

### 4.3.1 Materials for efficient SOT generation

The most studied SOT materials are the 5$d$ heavy metals Pt, Ta, and W. They exhibit efficiencies $|\xi_{DL}|$ ranging from 0.1 to 0.5 [2]. Alloys or multilayers containing 5$d$ metals, such as, e.g. AuPt or [Ti/Pt]$_n$, exhibit larger $\xi_{DL}$ compared to single-element systems due to enhanced electron scattering and resistivity [155–157]. 3$d$5$d$ metallic antiferromagnets such as MnPt and IrMn are also strong SOT sources, which additionally provide exchange bias to the FL [137,138,158–160]. Interface engineering by nitrification or oxidation of the heavy metal [161–163], insertion of ultrathin spacer layers [164,165], and use of electrically insulating antiferromagnetic layers [166] can further improve $\xi_{DL}$.

Among the 4$d$ metals, Pd presents sizable SOT in combination with PMA [167]. The 3$d$ metals that generate strong orbital currents can also be used as sources of SOT [49–52,56], as shown by recent demonstrations of current-induced switching in Cr/CoFeB and Zr/CoFeB bilayers [168,169], and 3$d$ ferromagnetic trilayers provide additional SOT sourcing mechanisms due to the generation of transverse spin currents [48,170].

Topological insulators (TIs) such as the bismuth chalcogenides Bi$_2$Se$_3$, (Bi$_{1-x}$Sb$_x$)$_2$Te$_3$, and BiSb have also drawn significant interest as they exhibit giant values of $\xi_{DL}$, typically 1-2 orders of magnitude larger than heavy metals [171–176]. The achievement of room temperature SOT-induced magnetization switching [172,173,175,177] confirms their high potential for SOT-MRAM. The fact that processing tools for physical vapor deposition, standardly used to deposit MTJ stacks, can be used to deposit TI layers with high SOT efficiency despite the polycrystalline nature and chemical disorder of the TI [178] is encouraging toward the realization of TI-based MTJs. Open questions remain concerning the presence of strong intermixing between ferromagnets and TI, damage due to the 400 °C heat treatment required by CMOS processing, and topological *vs* bulk origin of the spin current responsible for SOT [179]. Other materials that can be used to generate SOT are van der Waals materials [59,180,181] and oxide heterostructures [182]. For the time being, however, integrating these materials into devices for mass production remains a challenge.

### 4.3.2 SOT track conductivity

Together with $\xi_{DL}$, the conductivity of the SOT electrode determines the electrical power dissipated during switching. At the switching threshold, the power density is $P_{th} = j_{th}^2 \rho \propto \xi_{DL}^{-2} \rho$, where $\rho$ is the resistivity of the current channel and FL in parallel. The use of resistive materials can thus compromise the development efforts to introduce efficient SOT materials. Figure 7(a) shows how $\xi_{DL}$ correlates to $\rho$ and $P_{th}$ in different material classes. The typical FL used in the MRAM stack are made of CoFeB alloys with $\rho$ ranging from 80 to 130 $\mu\Omega$ cm. If the SOT track is too resistive, a significant portion of the current can be shunted into the FL [2,155,183]. In addition, the selector transistor is not only limited in current (affecting the transistor size/number and associated bit cell footprint), but also by voltage. This voltage strongly depends on technological node, ranging from 1.2 V in 40 nm nodes to 0.7 V beyond 22 nm nodes [184,185]. Hence, the SOT channel resistance has to be optimized, either by material choice or track design, in order to ensure sufficient current to operate the cell with the supplied voltage.

*4.3.3 MTJ structure*

The typical structure of modern MRAMs consists of a FL/MgO/RL/SAF stack [Fig. 7(b)], where the SAF is a synthetic antiferromagnet and the (storage) FL and the reference layer (RL) are made of CoFeB to ensure a high TMR ratio due to the spin-filtering effect in crystalline-MgO, reaching from 100 to 250% at room temperature in standard devices [23,24,128]. So far, no better materials compatible with industrial device requirements have been found to improve the TMR. Typical TMR reported in SOT-MTJ are on the order of 100% for out-of-plane MTJ [131], and 150% for in-plane MTJ [128]. This is mostly due to the top-pinned configuration of the stack, which seems to impact the TMR due to stress on the tunnel barrier, and the thin FL. However, thermal annealing up to 400 °C brings the TMR close to 150% [132,154]. In order to ensure that the RL has higher magnetic anisotropy than the FL, the RL is ferromagnetically pinned to a hard SAF layer by a thin nonmagnetic metal layer such as W or Ta. This metal insertion is equally important to decouple the hard layer texture, usually (111), to the (100) CoFe texture. The SAF relies on strong RKKY antiferromagnetic coupling between two or more ferromagnetic layers separated by thin nonmagnetic spacers [186]. The role of the SAF is to stabilize the RL and minimize the dipolar field of the RL radiated on the FL, which is crucial to ensure that the hysteresis loop of the FL is centered around zero-field [Fig. 7(c)]. Extensive stack engineering and optimization are required to reach the BEOL CMOS thermal budget, as the annealing increases the intermixing and induces degradation of various magnetic properties of the FL, RL, and SAF.

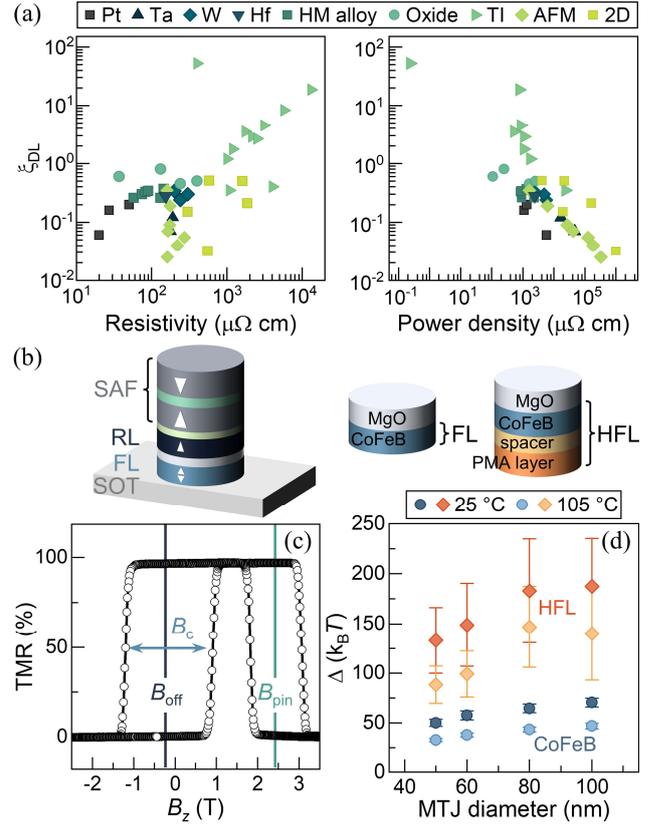

**Fig. 7.** (a) DLT efficiency of different material classes as a function of electrical resistivity and switching power density for different SOT layers comprising: heavy metals (HM), topological insulators (TI), antiferromagnets (AFM), and 2D materials. The power is estimated assuming a 0.9 nm thick CoFeB FL with a resistivity of 120 $\mu\Omega$ cm, and a 5 nm thick SOT layer for all materials. From: [2,155,157]. (b) Schematics of an MTJ pillar on the SOT track. (c) Easy-axis hysteresis loop of an MTJ device showing the FL reversal centered around the offset stray field produced by the SAF ($B_{off}$) and the RL reversal shifted by the pinning field of the SAF ($B_{pin}$). (d) Comparison of the thermal stability factors as a function of MTJ diameter for two realizations of the FL: standard CoFeB FL and hybrid (H)FL, adapted from [132].

*4.3.4 Free layer*

A crucial point of MRAM technology is the PMA of the FL, as perpendicularly magnetized stacks have better scalability and a smaller footprint than their in-plane magnetized counterparts. Interfacial PMA originates from the spin-orbit interaction and electronic hybridization effects between the FL and the SOT underlayer and/or MgO [187]. Magnetocrystalline anisotropy and strain can also contribute to improving the PMA. To circumvent data retention loss with scaling, STT-MTJ devices widely adopted the double MgO barrier concept [188,189] as in MgO/CoFeB/MgO to further increase interfacial anisotropy, or MgO/CoFeB/Ta/CoFeB/MgO to increase the magnetic volume by sandwiching two FL between the oxide barriers, providing thermal stability of the magnetization down to a cell diameter of ~20 nm.

We note that this approach is not straightforward for SOT, as MgO would be inserted in between the SOT and FL layers, resulting in degradation of spin current transfer. As an alternative, the concept of a hybrid FL, i.e., a FL composed of a reading CoFeB-based layer coupled to a magnetic layer or multilayer with stronger PMA [190,191] was recently demonstrated [132,192,193]. This solution significantly improves the thermal stability factor to $\Delta > 100$ in 50 nm pillars [Fig. 7(d)], while being compatible with the BEOL CMOS thermal budget and not affecting the TMR. In addition, this approach could benefit to introduce complex materials in the stack, such as TI, on which the growth of PMA is hampered by intermixing and roughness. Another alternative are ferrimagnets, as they possess strong bulk anisotropy and low saturation magnetization, in which the compensation of angular momentum reduces $j_{th}$ and improves the switching dynamics [194–197]. However, they are so far not compatible with MgO crystallization, which prevents reaching high TMR, and they typically degrade above 300 °C, making them impractical for memory applications.

## 5. Switching of 3-terminal MTJs

### 5.1 SOT switching and characteristics

Various metrics have to be established in order to assess the potential of SOT-MRAM for specific applications. The critical switching current, the write error rate (WER), or the endurance of the device are among the most important characteristics to report and combine with MTJ typical metrics (TMR, retention, anisotropy). This provides a benchmark to compare different SOT-MTJ systems, and against other technologies (see Table 3).

Whereas the typical switching characterization is performed by DC electrical measurements a few milliseconds after the writing pulse, the TMR offers a simple and efficient way to evaluate the outcome of the writing operation in scaled MTJs. Combined with time-resolved insight on the magnetization during and after the writing process, switching measurements provide information about the physical processes underlying the magnetization reversal, as discussed in the following.

The critical current $I_c$ is usually defined as the current pulse amplitude for which the switching probability ($P_{sw}$), shown in Fig. 8(a), is equal to 0.5. In SOT systems, the evolution of $I_c$ with the writing pulse length $\tau_p$ distinguishes two physical regimes of the magnetization reversal: an intrinsic regime in the high-amplitude short-pulse limit, and a thermally-activated regime in the low-amplitude long-pulse limit [76,199,200], as illustrated in Fig. 8(b). In the activated regime, the switching is assisted by thermal fluctuations and the current scales as $I_c \propto \log(\tau_p)$. This regime is characterized by wide dispersion of the switching onsets due to stochastic thermal assist. In the intrinsic regime, the critical switching current scales as $I_c = I_{c0} + q/\tau_p$, where $I_{c0}$ is defined as the intrinsic critical current and $q$ is an effective charge parameter that determines the rate at which angular momentum is transferred to the FL [199]. Such linear scaling [see the inset in Fig. 8(b)] was initially proven in 100-nm-wide Co dots patterned on Pt Hall bars, and later confirmed in SOT-MTJs based on Ta and W tracks, with diameter down to 50 nm [29,135,201]. Since the SOT switching of PMA systems requires an external in-plane field $B_x$, it is relevant to study $I_{c0}$ and $q$ as a function of $B_x$, as shown in Fig. 8(c). $I_{c0}$ is found to reduce monotonously with increasing $B_x$, as expected from Eqs. (5,6). New techniques to quantify the SOT efficiency at the SOT-MRAM cell level, and more refined analytical models capturing the correct switching mechanisms [163] are required to account for $B_x$ variation in bit-cell designs, especially when field-free switching solutions that involve a static magnetic field are implemented [131], or when considering perturbation of the device by an external field. Importantly, in order to extrapolate the current that is required to fulfill the specifications of an application, the switching measurements have to be extended to more events (typically $10^6$ events for cache applications). An important parameter is the write error rate WER = $\log(1 - P_{sw})$, which typically has a linear dependence on the writing current [see Fig. 8(d)]. The WER can also be used to identify infrequent errors, as observed in STT-MRAM due to backhopping and ballooning effects [202,203]. Unfortunately, there are currently few reports on this topic. The WER has been measured up to $10^5$ events [30,118,131,155,204] although with SOT-MRAM technology progressing rapidly, we expect more reports to appear in the coming years.

**Table 3.** Benchmark table of SOT-MTJ based on different SOT materials: W, W(O,N), Ta, TaN, Pt. The data are obtained from the same device design (MTJ diameter of 60 nm), and all of the switching properties were measured using a fixed in-plane field of 32 mT. The intrinsic critical switching current $I_{c0}$ is evaluated for a 50-nm-wide and 5-nm-thick SOT track an MTJ diameter of 50 nm using a parallel resistor model. $\Delta$ was extracted using the $B_c$ field distribution method [198]. Data: courtesy of IMEC.

| Material | W | W(O,N) | Ta | Pt | TaN |
|---|---|---|---|---|---|
| Anneal (°C) | 300–375 | 300 | 300–375 | 300–400 | 300-400 |
| $\rho_{SOT}$ (μΩ cm) | 140 | 160 | 180 | 32 | 240 |
| TMR (%) | 110 | 80 | 120 | 125 | 120 |
| FL type | CoFeB | CoFeB | CoFeB | hybrid FL | CoFeB |
| $B_c$ (mT) | 95 | 75 | 120 | 125 | 160 |
| $B_k$ (mT) | 270 | 220 | 270 | 240 | 350 |
| $\Delta$ ($k_BT$) | 50 | 40 | 58 | 150 | 59 |
| $j_{c0}$ (MA/cm²) effective SOT+FL | 116 | 93 | 140 | 120 | 118 |
| $I_{c0}$ (μA) @ 50 nm (with 10 nm overlay) | 410 (480) | 340 (394) | 500 (570) | 440 (500) | 440 (500) |
| RA (Ω μm²) | 10-5000 | | | | |

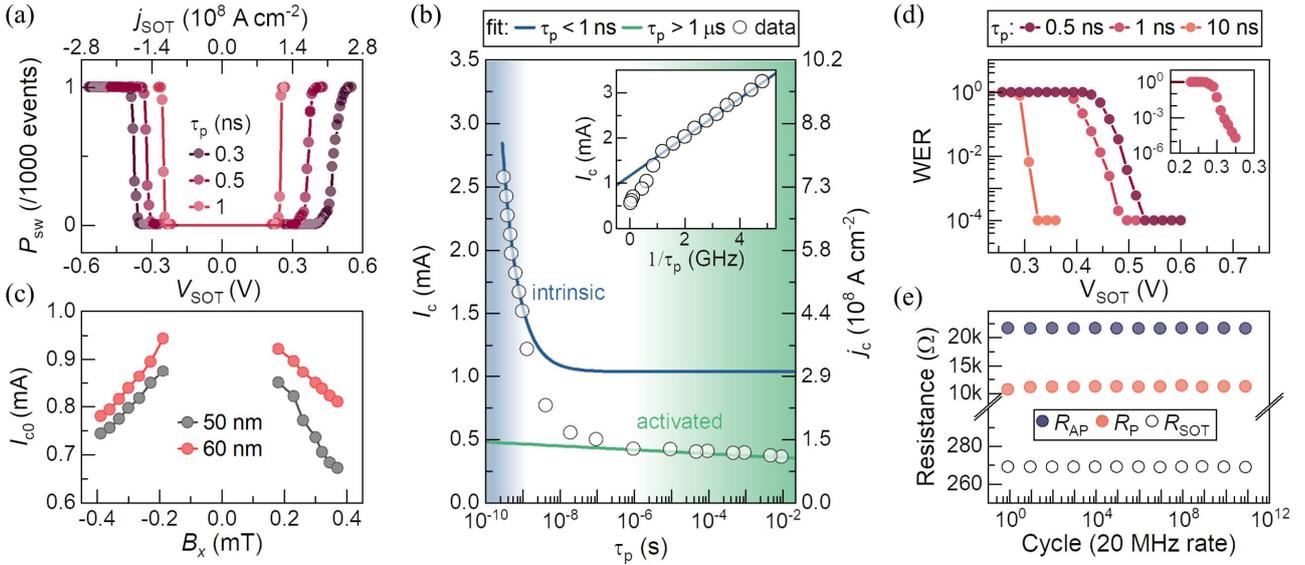

**Fig. 8.** (a) Switching probability as a function of SOT bias voltage ($V_{SOT}$) for different pulse widths ($\tau_p$), adapted from [131]. (b) Critical switching current as a function of the pulse width $\tau_p$, adapted from [76]. (c) Intrinsic critical current as a function of assisting in-plane field $B_x$. (d) Write-error rate of $10^4$ events for different pulse widths, inset: write-error rate of $10^6$ events for 10-ns-long pulses. (e) Endurance test of $10^{12}$ cycles at $j_{SOT}$ = 140 MA cm$^2$ and $\tau_p$ = 0.3 ns, adapted from [131]. Data panel (c), (d): courtesy of Imec and Hprobe [205].

Finally, the endurance test overpassing $10^{12}$ events has also shown that SOT-MTJs are very resilient to degradation and electromigration [Fig. 8(e)], despite the large current densities in the SOT track and severe writing stress imposed by the current.

### 5.2 Real-time detection of MTJ switching

Besides the standard post-pulse readout, all-electrical measurement schemes can be used to probe the time evolution of individual switching events in an MTJ. The real-time readout of the TMR was first employed during the writing operation of STT-MTJ cells [206–208]. More recently, measurement schemes adapted to probe the SOT-driven dynamics in 3-terminal MTJs were employed [31,109,209]. These single-shot electrical measurements allow for the investigation of the magnetization dynamics at the device level, including stochastic aspects and switching distributions. Time-resolved measurements provided the first insight into the latency and actual speed of SOT-induced magnetization reversal [31,135,210,211], the dynamic interplay of SOT, STT, VCMA, and self-heating [109], and the origin of writing errors known in the quasi-static picture as backhopping [203].

The timescale relevant for the writing operation of the SOT-MRAM is 0.1–10 ns, crossing the borderline between the intrinsic and thermally-activated regimes [76,135]. The switching of sub-100 nm MTJs by pulses in this time window showed that the magnetization reverses via a single jump from the initial state to the final state without intermediate levels [31]. The reversal is completed within the duration of the current pulses and is preceded by a delay [see Fig. 9(a)], similarly to the switching of perpendicular MTJs by STT [206,212]. Such incubation delay is a typical feature of STT switching due to the initially collinear alignment of the magnetizations ($\mathbf{m} \parallel \mathbf{m}_{RL}$) discussed in Section 2 [208,213]. However, it is an unexpected feature in the case of SOT because of the initial orthogonal alignment of DLT and free layer magnetization ($\mathbf{m} \perp \boldsymbol{\sigma}$). Nevertheless, the observation of up to several ns-long delay times ($t_0$) in MTJ devices with a ferromagnetic FL [31], as well as in ferrimagnetic dots [110], indicates that the delayed reversal is a common characteristic of SOT switching for currents close to $I_c$. The delay can be suppressed by increasing the current above $I_c$ and is therefore not an intrinsic feature of SOT switching. It is in fact most pronounced for the longer and low-amplitude current pulses, indicating its relation to thermal activation. As the typical FL used in SOT-MTJs have strong PMA, very large SOT are required to nucleate a domain and initiate the switching process. During the SOT pulse, the temperature of the FL increases and thus gradually reduces its PMA, lowering the switching energy barrier until the SOT are large enough to initiate the reversal [31,109]. In this way, the critical switching current is lower than expected in the intrinsic regime, however, at the expense of finite latency. This activation time and the role of heating thus differ from the incubation delay in STT switching, where heating enhances the torque by increasing the transversal component of the magnetization in the FL.

From the point of view of applications, latency and jitter are undesirable features as they limit the maximum frequency of operation. The current flowing in the SOT track is determined by the SOT bias $V_{SOT}$. Since the switching time, as well as $t_0$, scale as $\exp(-V_{SOT})$ or $1/V_{SOT}$ – in the thermally-activated or intrinsic regime, respectively – the most straightforward way to accelerate the switching is to increase the SOT bias. Due to the non-linearity of the scaling, even a small increase of $V_{SOT}$ has a significant effect on $t_0$, as seen in Figs. 9(a) and (b) where increasing $V_{SOT}$ by 25% (0.1 V) reduces the average delay four times. The fastest reproducible switching of SOT-MTJ in real time was acquired using <0.3 ns-long pulses [31], and faster switching appears possible [107]. The increase in $V_{SOT}$ is not the only option to accelerate the switching: Figures 9(c) and (d) show that an increase of $B_x$ or the application of a voltage pulse across the MTJ (due to STT and VCMA assist) efficiently minimize the incubation delay ($t_0$) and the duration of the reversal phase ($\Delta t$). Importantly for device operation, the dispersion of these parameters recorded over long sequences of single-shot switching events also narrows down significantly. It remains an open question to what extent these observations will remain valid in MTJs scaled down to the limit of macrospin switching [214].

## 5.3 Complementary approaches to SOT switching

The SOT geometry and the use of 3-terminal MTJ devices allow for writing schemes that take advantage of multiple switching mechanisms, as schematized in Fig. 10(a). We discuss them in the following.

### 5.3.1 SOT-STT switching

Switching of an MTJ by a combination of SOT and STT biases was initially proposed to lower the critical writing current and energy of the STT-MTJ [215,216]. By incorporating a strong-SOC material into the STT cell, it is possible to use the 2-terminal geometry suitable for footprint reduction [116], as well as to lift the requirement of an external magnetic field [122,125,149,154,211]. In this case, however, STT serves as the main switching mechanism, whereas the SOT supplies the in-plane spin polarization to assist the reversal. It is important to note that the relative strength of STT and SOT applied to the FL depends on the partition of the current flowing in the SOT track and across the MTJ, which changes with both $V_{SOT}$ and $V_{MTJ}$, namely the bias applied to the SOT track and top electrode of the MTJ, respectively [31]. Overall, SOT-STT switching realized using materials with enhanced charge-to-spin conversion efficiency [2,217] represents a promising pathway towards the realization of high-density MRAMs.

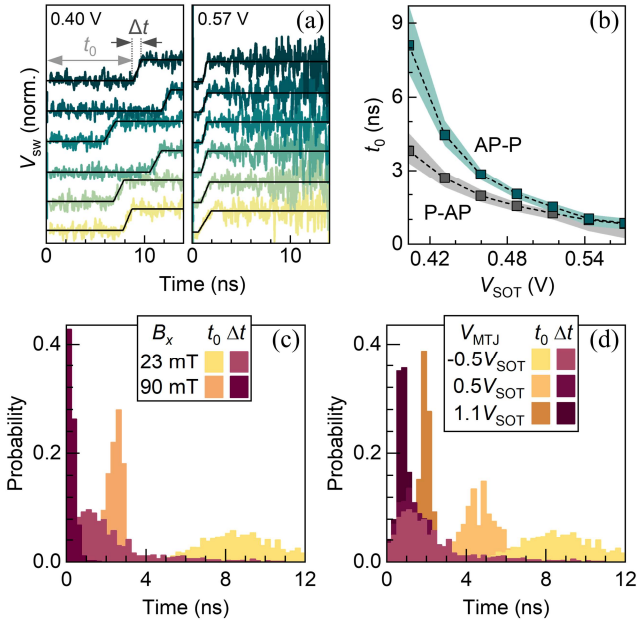

**Fig. 9.** (a) Switching time traces of individual AP-P switching events showing a reduction of the incubation delay ($t_0$) and the reversal time ($\Delta t$) with an increase of the SOT pulse $|V_{SOT}|$ from 0.4 V to 0.57 V and (b) The delay $t_0$ obtained from 1000 switching trials for both switching polarities in the same 80-nm MTJ device. (c,d) Statistical distribution of $t_0$ and $\Delta t$ for different (c) $B_x$ and (d) voltage across the MTJ ($V_{MTJ}$). Reproduced from [31,135].

## 5.3.2 VGSOT switching

The voltage gate-assisted SOT (VGSOT) is a combinatory approach to MTJ switching that exploits the VCMA effect to reduce the switching energy. This has been found to be particularly effective to enable efficient and ultrafast SOT switching applications [31,109,118]. The VCMA refers to the change of the interfacial PMA caused by voltage-induced charge transfer at the FL interface [218,219] [see Fig. 10(a)]. Since the effect is electrostatic and takes place without significant current flow across the junction, it can be extremely energy efficient. In principle, VCMA alone can be employed for the switching of MTJs, however, the VCMA efficiency of the CoFeB/MgO system is not strong enough to be competitive [118,220–223] and thus other materials are needed [224]. In the VGSOT scheme, the SOT induces switching of the FL while the bias applied to the top MTJ electrode ($V_{MTJ}$) lowers the reversal energy barrier through the VCMA effect. This is advantageous in several ways. First, since the switching threshold is proportional to the PMA of the FL, applying $V_{MTJ}$ allows for significantly reducing $V_{SOT}$ [see Fig. 10(b)]. A reduction of the SOT current by 25–38% per $V_{MTJ}$ = 1 V [118,225] or a reduction of the writing energy by 45% at 1 V [118] have been achieved for the same switching time. Second, lowering the PMA significantly reduces the delay time $t_0$. Third, $V_{MTJ}$ can be used also to increase the PMA during the read operation to limit read disturbances. Fourth, $V_{MTJ}$ can make the switching threshold more symmetric. Due to the hard layers' stray field, $V_{SOT}$ is generally asymmetric for P-AP and AP-P switching, which implies that the maximum frequency of operation is limited by the less efficient of them. Whereas the stray field asymmetry cannot be fully removed, the related voltage imbalance can be corrected by $V_{MTJ}$ [see Fig. 10(c)]. Finally, $V_{MTJ}$ can serve as an MTJ selector, thus enabling a multi-pillar cell structure as sketched in Fig. 10(d), which can effectively reduce the cell size to address the density limitation of SOT technologies [118].

## 5.3.3 Combined effects of SOT, STT, VCMA and self-heating

The combined impact of SOT and STT biases ($V_{SOT}$ and $V_{MTJ}$) promises higher switching efficiency than either of the two effects separately [109,112,211]. In general, $V_{MTJ}$ can induce an STT current or the VCMA effect – or both – depending on the resistance-area (RA) product of the barrier. Moreover, the additional current across the MTJ also increases the heating load of the pillar, which further reduces the magnetic anisotropy of the FL. Thus a single SOT-MTJ cell can take advantage of VCMA, STT, and self-heat simultaneously [109,226]. The scaling and symmetries of these effects are not the same, which can be leveraged to tailor the MTJ cell performance. Notably, it has been reported in CoFeB/MgO devices with low RA, that this combinatorial assist to SOT offers a reduction of the writing energy by 40% and can even surpass the energy-efficiency of STT without compromising the writing speed [see Fig. 10(e)] [31,109].

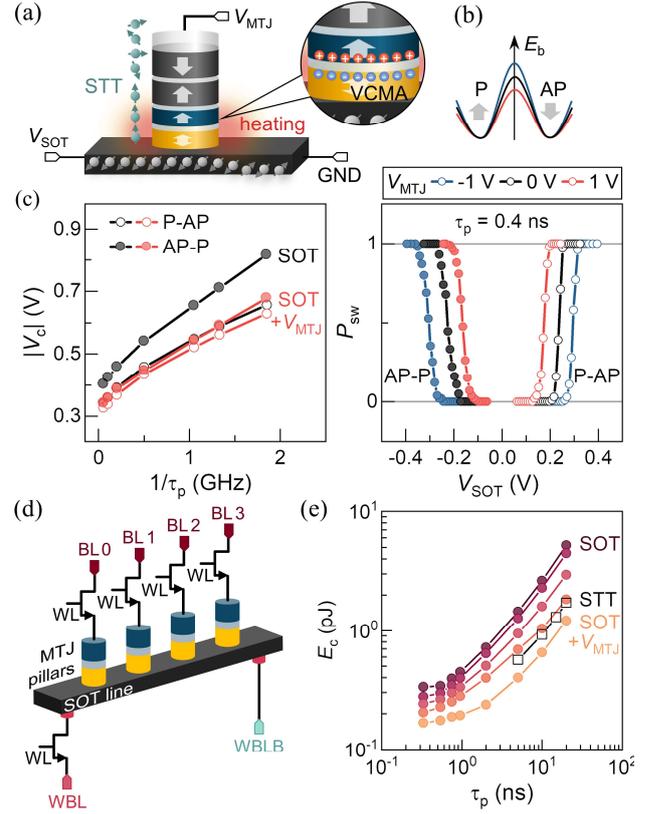

**Fig. 10.** (a) Illustration of the different effects that assist SOT switching upon application of $V_{MTJ}$: VCMA, STT, and self-heating. A positive bias induces electron accumulation at the FL/MgO interface, which lowers the FL anisotropy and reduces the switching threshold. (b) Switching probability as a function of $V_{SOT}$ and $V_{MTJ}$, reproduced from [118]. (c) Use of $V_{MTJ}$ to compensate the threshold asymmetry of SOT switching, reproduced from [135]. (d) Multi-pillar VGSOT cell structure with four MTJs sharing the same SOT track, where $V_{MTJ}$ serves as the MTJ selector. (e) Critical switching energy for different pulse widths for SOT, STT, and combined SOT-STT switching, reproduced from [109].

Because Joule heating is proportional to the power dissipated in the MTJ pillar, $V_{MTJ}^2/R_{MTJ}$, whereas both VCMA and STT are proportional to $V_{MTJ}$, but the VCMA depends on sign($V_{MTJ}$), whereas STT depends on sign($V_{SOT}V_{MTJ}$), these effects can be disentangled from each other and the critical SOT voltage as a function of $V_{MTJ}$ can be expressed as [109]

$$V_c(V_{MTJ}) = V_c^{SOT}\left[1 - \left(\frac{2\varepsilon V_{MTJ}}{M_S t_{FL} t_{MgO}} \pm \frac{\eta V_{MTJ}}{RA(1-b|V_{MTJ}|)} + \frac{\zeta V_{MTJ}^2}{RA(1-b|V_{MTJ}|)}\right) \cdot (B_K \mp B_{off})^{-1}\right], \quad (8)$$

where $b$ is a coefficient used to approximate the $V_{MTJ}$ dependence of the TMR and $\varepsilon$, $\eta$, and $\zeta$ are the VCMA coefficient, STT efficiency, and self-heating coefficient, respectively. Moreover, VCMA, STT, and self-heating affect the switching onset differently as the latter is most pronounced on timescales longer than a few ns and devices larger than about 100 nm, whereas the VCMA dominates at sub-ns timescale and in small MTJ. This suggests that the role of VCMA becomes increasingly important in downscaled devices.

## 6. SOT integration and system-level considerations

### 6.1 SOT-MTJ integration and challenges

To cover a broader range of envisioned applications, further development of cost-effective and versatile MRAM technology will be crucial. It will also be necessary to ensure proper downscaling of the MTJ diameter and cell-to-cell pitch to maintain competitiveness at advanced technology nodes. As a rule of thumb, because of the required analog circuitry, it is desirable to keep the MRAM bit cell about three times smaller than the SRAM cell size at a given technology node, so that the benefit is worth the technology change. In terms of fabrication flow, SOT requires two additional modules compared to STT flow, such as forming and aligning the SOT channel. The main integration process steps consist of four modules [131], as illustrated in Figs. 11(a) and (b): i) Patterning, metal filling and planarization of W vias to contact the SOT track followed by the deposition of the MTJ stack, ii) MTJ pillar definition using advanced lithography, e.g., 193 nm immersion, followed by ion-beam etching and encapsulation, iii) SOT track patterning, completed with ion beam etching and followed by encapsulation, and iv) a Cu dual-damascene process for routing back to selector transistor (through M1 to M4 metal levels) or to interconnect the top and bottom electrodes (BE) routing to pads for single-cell device testing. Each of these modules is typically followed by an oxide refill and chemical-mechanical planarization (CMP) to flatten the surface at the nm level. A typical single SOT-MTJ cross section resulting from such integration scheme is presented in Figs. 11(c)-(e).

Given that SOT-MRAMs share the same technological platform as STT-MRAMs, SOT benefit from the processes implemented in STT-MRAM foundries to improve device yield and performance: high quality and homogeneous stack growth, morphology, and etch impact on magnetic properties and TMR, diffusion barriers to reach BEOL thermal budget (400 °C). However, in STT, the FL is on top of the MTJ stack (bottom-pinned) whereas, for SOT, it is at the bottom (top-pinned). This change of stack forces to re-design material growth protocols, and introduce specific seed layers to achieve the desired crystallinity of the MTJs.

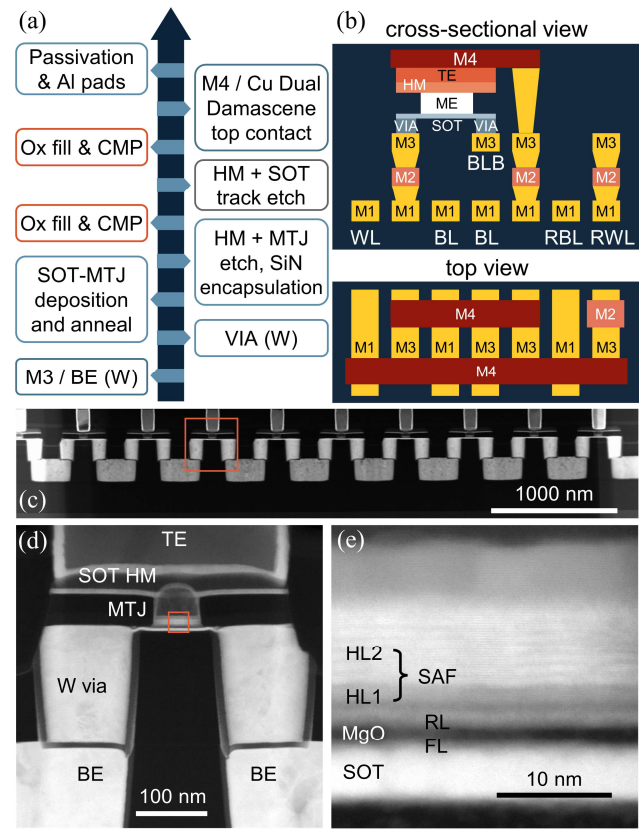

**Fig. 11.** Integration workflow resuming the main modules required to realize SOT-MTJ MRAMs. BE (bottom electrode), CMP (chemical mechanical polishing), IBE (ion-beam etching). In case of integration with CMOS, BE (TE) are replaced by M3 (M4) metal level modules. (b) Layout and cross-sectional illustration of the traditional 5-Terminal SOT cell. (c) High-angle annular dark-field transmission electron microscopy image of a SOT-MRAM chain, (d) zoom on a cell showing a hybrid FL-MTJ with a straight profile placed on a continuous SOT (Pt) layer and contacting two smooth W-vias, and (e) zoom on the MTJ pillar showing sidewalls clean of shorts across the MgO barrier. Image: courtesy of IMEC, adapted from [131,132].

In particular, achieving top-pinned structures compatible with BEOL post thermal annealing at ~400 °C has been a challenge, especially for MTJ with PMA. In fact, the atomic intermixing at interfaces and strain induced by the growth of differently textured materials lead to an overall device performance deterioration, for example, FL and SAF loss of PMA (and $M_s$). Whereas in-plane MTJs were readily BEOL-compatible owing to the thicker FM layer, perpendicularly magnetized MTJ stacks reached 400 °C compatibility only two years ago [190,227], and SOT-MTJ stacks were recently demonstrated [132,154].

A challenge that can significantly impact the yield is that the etching of the MTJ has to stop on the thin SOT channel precisely (typically less than 5 nm thick) without degrading its conductivity or causing vertical shorts due to metal re-deposition on the tunnel barrier sidewalls. Various strategies are envisioned to calibrate the etch conditions (on MgO, in MgO, in SOT,…) [154,201,228]. Pitch scaling, targeting a separation <100 nm of MTJs with diameter ~30 nm, and the fabrication of multi-pillar devices on the same SOT track [Fig. 10(d)] [118,149,229] will however increase shadowing and likely impact yield by increasing the short failure rate induced by sidewall metallic re-deposition during etching, calling for optimized SOT-MTJ etch module development.

Regarding the SOT track module, the major challenge is the alignment precision (ideally self-alignment) and the width minimization with respect to the MTJ diameter, to maximize write efficiency and minimize total charge current, while over-etch in oxide should limit vertical shorts.

### 6.2 SOT cell dimension optimization

The challenges currently hindering large-scale SOT technology adoption in the embedded domain are primarily linked to the writing operation, memory density, and the integration of the SOT stack with standard logic technology. To properly dimension a cell, one must account for the targeted technology node, which will impose the maximum dimensions of SOT track width and length compatible with access via pitch and selector (transistor) footprint. For example, a SOT track width of <50 nm and length of ~100 nm should be compatible with sub-22 nm nodes.

On the other hand, such dimensions are not yet studied or reported, so that the cell properties are extrapolated from existing devices. This is done using compact models that capture the key switching mechanisms and physical properties of the devices, which are deduced from either physics-based or behavioral models [183,230].

Figure 12(a) illustrates the extrapolations derived from W-based SOT MRAMs with 160 nm and 100 nm channel width for a 32 nm device. The 160 nm channel width writing properties correctly extrapolate to a 100 nm device. Importantly the critical current at 1 ns reduces to 100 µA for a 32 nm track according to SOT compact models [231]. As this is still insufficient to match the sub-100 µA requirement, corresponding to advance single transistor typical maximum current delivery capability, one can adapt the model to account for the improved charge-to-spin conversion ratio assuming more progress on the material side. This is highlighted in Fig. 12(b), where the critical switching current at 1 ns is presented as a function of $\xi_{DL}$. Based on such reasoning, $\xi_{DL} \sim 1$ would be required to match the technology requirements.

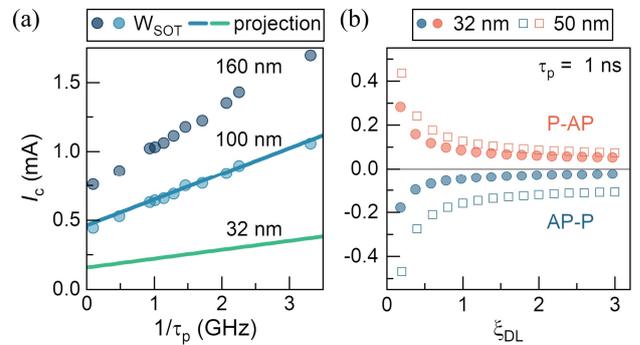

**Fig. 12.** SOT critical switching current for different SOT-MTJ device dimensions. (a) $I_c$ measured in devices with 160-nm- and 100-nm-wide W-SOT tracks and 60-nm-wide MTJ (symbols), and associated extrapolation (lines) to narrower SOT tracks based on behavioral compact model [231]. (b) $I_c$ for P-AP and AP-P switching for different values of the SOT efficiency $\xi_{DL}$.

### 6.3 Bit-cell configurations and design technology co-optimization

Thorough design technology co-optimization and accurate technology assumptions are important to estimate the cell footprint for the design of embedded memories. When benchmarking against SRAM, one should consider that SRAM is designed with front-end-of-line (FEOL) design rules requirements (usually tightest pitch at M1), whereas the SOT cell height is mostly determined by BEOL patterning design rules which set constraints in terms of via pitch (usually M2 to M4 level).

Although commonly denominated as a 3-terminal device, a regular SOT-MRAM bit-cell is, in reality, a 5-terminal device, i.e., the standard 3-terminal SOT-MTJ device and the gate of two selector transistors [labeled SOT-5T, shown in Fig. 13(a)]. Consequently, the SOT-5T, utilizes two selectors with read and write (RD/WR) word-lines (WLs), bit-lines (BLs), and WR bit-line bar (BLB). Routing these five signals restricts the cell scaling from the BEOL point of view. Moreover, current requirements, and separate RD/WR selectors, are placing limitations on the bit-cell area from the FEOL point of view (metal pitch), which must accommodate a sufficient amount of Fins or a large-enough transistor to deliver the required writing current. In addition, one must also account for BL/BLB series resistance adding up to the SOT track. This can have a severe impact on the ability to drive enough current through the SOT track without increasing the FEOL overhead. The BL and BLB resistances are extremely dependent on the targeted technological node. Consequently, one will have to consider design strategies with optimal high-performance (HP) but increased FEOL, or with degraded performances but optimal density: high-density (HD).

Keeping in mind these limitations, three alternative bit-cell solutions with four terminals were proposed [231]: shared WL (SOT-SWL) of types 1 and 2, shown in Fig. 13(b) and (c), and shared BL (SOT-SBL) shown in Fig. 13(d). The configurations with shared WL have the benefit to solving the inherent voltage/current writing imbalance required for the P-AP and the AP-P transitions, induced by the SAF stray field. Indeed, a writing asymmetry implies that the maximum frequency of operation for the cell is limited by the least favored transition. This approach can leverage a $V_{MTJ}$ assist (by STT or VCMA), which can be applied to the top electrode, as this assist will either increase/decrease the SOT switching voltage (see Section 5).

From this point onwards, additional integration strategies can be applied to minimize the area of the bit-cell, signal parasitics (resistance, capacitance), and the ability to drive the SOT current for SOT-HP and SOT-HD solutions, as detailed in [231]. Figure 13(e) shows several examples of SOT bit-cell areas benchmarked against SRAM for different design strategies at 5 nm, while Figure 13(e) evaluates for selected HP and HD designs the scaling of the bit-cell area for different technological nodes.

### 6.4 Macro-design architecture and PPA analysis

In this section, we define important challenges from a design-architecture point of view to tackle the adoption of SOT as the embedded memory of choice (compared to SRAM and STT-MRAM) using a detailed power, performance, and area (PPA) analysis.

The first step to build the PPA analysis is to enable the development of an accurate SOT-MRAM process design kit (PDK) containing the compact model (usually coded in VerilogA), layer definitions for the appropriate technology node, parameter, and port definitions. These models convert the SOT-MRAM cell into electrical parameters and can be used with existing similar CMOS library models in SPICE (integrated circuit simulators) for time-domain analysis of memory macro performance [232]. For the device models, various alternatives exist with the constraint of having a sufficiently fast simulation time response of a memory array macro (containing typically > 250 kbit-cells) with the accuracy of the model. Purely physics-based models relying on simplified physics (macrospin here) generally fail in capturing nano-dimensions complex mechanisms, whereas advanced models like micromagnetic simulations are usually very time-consuming. Thus, a hybrid model that combines the speed of analytical fitting-based models (behavioral) obtained from experimental data and the accuracy of physics-based models is best suited to enable memory macro design [230,233–235].

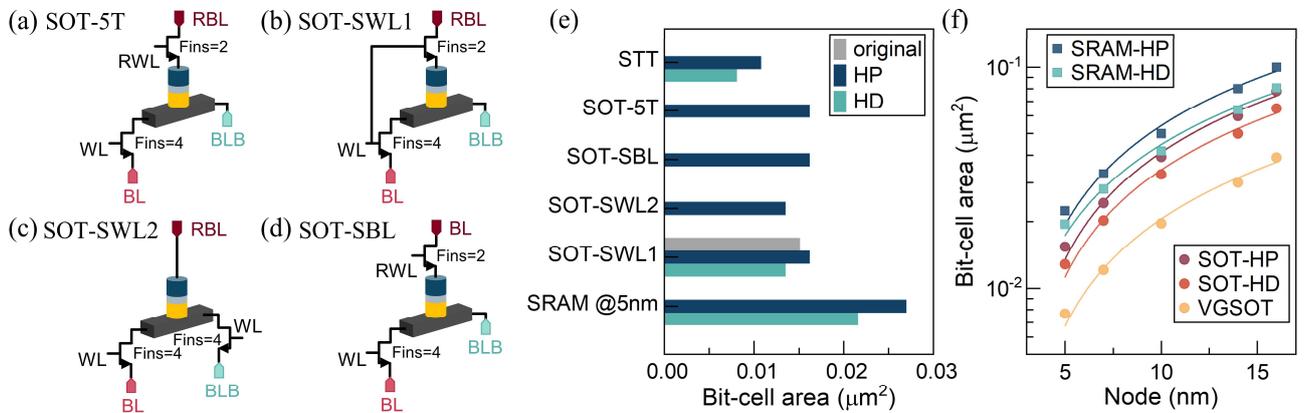

**Fig. 13.** Different SOT bit-cell configurations with a transistor selector: (a) Traditional 5-terminal SOT (SOT-5T), 4-terminal SOT (b) shared WL type 1 (SOT-SWL1) and (c) type 2 (SOT-SWL2), and (d) shared BL (SOT-SBL). (e) SOT-MRAM bit-cell area compared to SRAM and STT-MRAM and (f) scaling of the bit-cell area for different technological nodes. SOT-MRAM cell height is governed by BEOL patterning compared to SRAM cell.

Once the hybrid compact model has been calibrated via real device measurements, an additional model layer is required to capture statistics, variability and the process corners (the extreme variation of fabrication parameters). Combined with the design technology co-optimization strategies described in the previous section for various bit-cells designs, one can determine the RD/WR margins, the supply voltage(s), and required peripheral circuitry for a specific SOT-MRAM array size. The periphery includes local input/output (I/O) terminals, sense amplifiers, write drivers, pre- and post-decode, and the local control. However, application domain performance will impose partition of the main macro in order to mitigate the parasitics, and the related delay constraints to access the SOT memory cell. This can be done using a butterfly-based design architecture, as shown in Fig. 14(a). The sub-array banks [see Fig. 14(b)] populating each butterfly wing, and their subsequent stitching together is obtained via local repeaters and local I/Os for top and bottom arrays. Their amount (N) will also depend on the parasitics, delay and performance constraints.

Assuming such macro design, one can perform a PPA analysis for various SOT-MRAM designs (5T, 4T-HD, 4T-HP, VGSOT) whose cell layouts were optimized for the embedded domain at the 5 nm technology node. The macro has a capacity of 128 kbits and is benchmarked against other technologies (SRAM, STT, VCMA), as summarized in Table 4.

The maximum bank size is 32 kbits (128 rows x 256 columns). The number of decode levels increases for larger memory sizes to accommodate performance requirements at the cost of cell area efficiency (ratio of memory cells to peripheral circuits). A single-ended sensing scheme is considered for probing the MTJ state. Whereas the power consumption and access delay for the macro-level specifications of memory are determined by technology, these specifications for a memory sub-system (like a cache) in the system on chip are heavily application-dependent. SOT-MRAM read and write timings compare favorably against other embedded magnetic memory options but at the cost of higher power per bit. The reduced current through the MTJ stack also helps to push the endurance for SOT-based variants of embedded magnetic memories to ~$10^{14}$. The VGSOT memory cell configuration allows for the reduced bit-cell area (per bit) based on the number of MTJ pillars on the SOT track. The high resistance required for effective VCMA operation leads to slower read access times for VCMA and VGSOT memories.

SOT-MRAM compares favorably against SRAM due to its non-volatility and thus considerable energy savings that can be achieved by turning off memory instances/banks based on the application profile and data placement policies. Based on previous assumptions, one can map the potential of embedded MRAM technology solutions based on the application domain (e.g., Internet-of-Thing, Mobile, High Performance) by carrying a cross-layer workflow, as shown in Fig. 14(c). This projection highlights that SOT-MRAM presents a competitive option for the higher and last level embedded caches in several domains by adjusting design, and improving material parameters. We note that the specifications required for L1–L4 levels are different for each application domain. In particular, VGSOT can improve significantly density and writing at the only compromise of reading speed, whereas standard SOT enables closer to core higher performance options at the cost of density.

**Table 4.** Comparative table of potential embedded memory solutions at the macro level (5 nm node, 128 kbit) based on IMEC internal evaluations.

| Specs | Embedded memory options | | | | | | | | | |
|---|---|---|---|---|---|---|---|---|---|---|
| | SRAM | | STT | | SOT | | | VCMA | VGSOT | |
| | HP (122) | HD (111) | HP (1.5 CPP) | HD (2 CPP) | 5T | HP (4T) | HD (4T) | | 2MTJ | 4MTJ |
| Bit-cell area (μm$^2$) | 0.028 | 0.021 | 0.0108 | 0.0081 | 0.0162 | 0.0162 | 0.0135 | 0.0208 | 0.0122 | 0.009 |
| RD power/bit (nW) | 18.7 | 7.28 | 7.44 | 6.20 | 25.5 | 29.0 | 32.9 | 3.16 | 3.16 | 2.96 |
| WR power/bit (nW) | 25.7 | 9.85 | 29.2 | 24.4 | 31.4 | 34.7 | 35.3 | 59.2 | 47.5 | 46.5 |
| RD latency (ns) | ~0.80 | ~1.50 | ~2.89 | ~3.75 | ~1.00 | ~1.00 | ~1.00 | ~10.0 | ~10.00 | ~10.00 |
| WR latency (ns) | ~0.80 | ~1.50 | ~7.78 | ~20.00 | ~2.00 | ~1.40 | ~2.00 | ~1.00 | ~1.00 | ~1.00 |
| Endurance | $10^{16}$ | $10^{16}$ | $10^7$ | $10^9$ | $10^{14}$ | $10^{14}$ | $10^{14}$ | $10^{14}$ | $10^{14}$ | $10^{14}$ |
| $V_{DD}$ (V) | 0.7 | 0.7 | 0.7 | 0.7 | 0.7 | 0.7 | 0.7 | 1.4 | 0.7 | 0.7 |

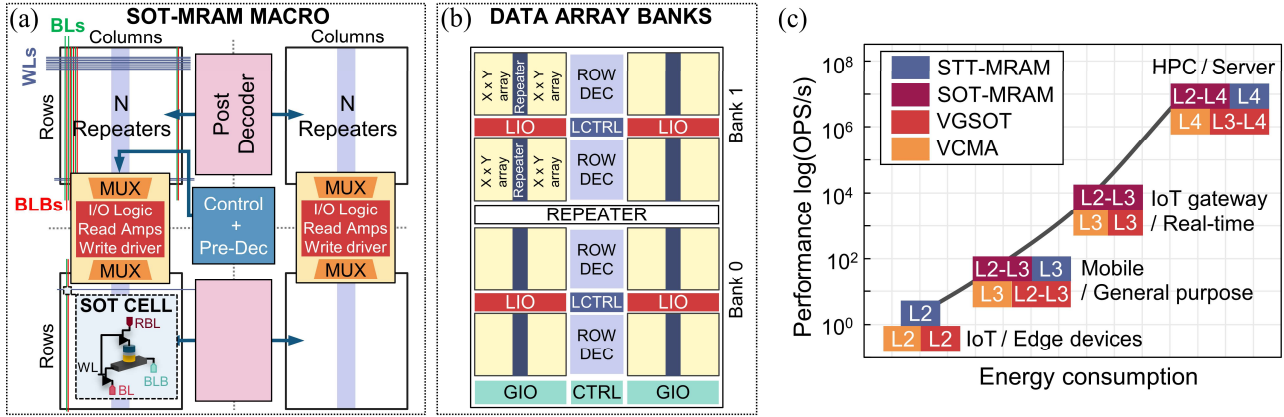

**Fig. 14.** (a) SOT-MRAM memory macro butterfly design for embedded last level caches including main peripherals: multiplexer (MUX), I/O logic including sense amplifiers and write drivers, pre/post-decode controllers. (b) Memory macro data array banks including local periphery and repeaters to minimize parasitics and delay constraints to access the memory. (c) Extrapolated operations per second to estimate memory technology feasibility based on target applications.

## 7. Perspectives beyond memory applications and conclusion

The previous section showed that SOT-MRAM has great potential as an embedded replacement for SRAM in various application domains that run close to the clock speed of the central processing unit. Meanwhile, the rapid growth of data volume and applications operating on large datasets, such as artificial intelligence and cloud computing, poses a major challenge for modern computation systems, calling for the development of unconventional computing architectures. In the traditional von Neumann architecture, the (volatile) processing unit and the (non-volatile) memory unit are physically separate, and data need to be frequently transferred between these two. Thus, a significant portion of the energy and computation time is consumed by this transfer – a problem known as the "von Neumann bottleneck" or "memory wall". Introducing non-volatility in the cache memory with increased density will partly help in limiting data transfer out of cache, but also offers the possibility to move towards more and more local data processing for reasons of performance, energy savings, security, and reliability. The general idea is to bring the calculation closer to the data to be processed, by moving the calculation to the peripheral circuits of the memories (near-memory computing), or even to integrate the calculation directly into the memory, i.e., in-memory computing (IMC), to significantly reduce the number of accesses to the memory during calculation operations. Non-volatile resistive memory elements will play a key role in their development [11,236]. In this context, MRAMs present the unique asset of being fast, dense and CMOS compatible, fulfilling many IMC requirements. SOT-based devices further include a third terminal, which allows for tuning separately the reading and writing paths for neuromorphic computing and logic applications, but also adds a tunable communication channel of interest for bio-inspired and Ising computation schemes [237].

In particular, deep neuromorphic networks and inference concepts are nowadays heavily used for pattern recognition tasks. From the hardware point of view, the ability to program very high resistance non-volatile elements (weight) with tight variability and no compromise on the write process is mandatory to perform vector matrix multiplication. Preliminary experimental and design works show that VGSOT-MRAM is indeed an excellent candidate to implement multilevel cell weights for inference accelerators running quantized deep neural networks [238,239]. Similar concepts can be extended when using DW-based devices where the DW is nucleated by STT and moved by SOT [240–242]. All-electric IMC logic devices based on magnetic DWs or skyrmions are also possible [96,240,242–246]. Stochastic computing is another area where the third terminal of SOT devices enables different circuit designs dedicated to solving optimization problems and to perform invertible logics [21,247–249].

In conclusion, SOT-induced magnetization switching and related technologies are only about a decade old and have progressed rapidly in terms of device proofs-of-concept and material innovation, making it closer to industrial applications. We have covered in this review the fundamentals of SOT-driven magnetization reversal in devices, focusing on nanoscale SOT-MTJ cells. We discussed the main figures of merit and the challenges that must be addressed to meet different memory specifications in terms of speed, power, materials and stack development, technology integration, and macro circuit considerations. Our survey shows that SOT-MRAM is a promising competitor of SRAM embedded memories and that it has a great potential to be used in a wider range of applications, ranging from IMC to neuromorphic and stochastic computing schemes. For this to occur, application-oriented developments at the single device level should be conducted in close collaboration with circuit designers in order to bridge device and system-level requirements.

## List of abbreviations

| | |
|---|---|
| AFM | antiferromagnet |
| AP, P | antiparallel, parallel |
| BEOL, FEOL | back-end-of-line, front-end-of-line |
| BL, WL | bit-line, word-line |
| BLB | bit-line bar |
| CMOS | complementary metal-oxide-semiconductor |
| CMP | chemical-mechanical planarization |
| DLT, FLT | dampinglike torque, fieldlike torque |
| DMI | Dzyaloshinskii-Moriya interaction |
| DW | domain wall |
| FL | free layer |
| HD, HP | high-density, high-performance |
| IMC | in-memory computing |
| MOKE | magneto-optical Kerr effect |
| MRAM | magnetoresistive random access memory |
| PDK | process design kit |
| PMA | perpendicular magnetic anisotropy |
| PPA | power, performance and area |
| RA | resistance-area |
| RL | reference layer |
| SAF | synthetic antiferromagnetic structure |
| SOI | spin-orbit interaction |
| SOT | spin-orbit torque |
| STT | spin transfer torque |
| TMR | tunnel magnetoresistance |
| TI | topological insulator |
| VCMA | voltage control of magnetic anisotropy |
| VGSOT | voltage gate-assisted spin-orbit torque |
| WER | write error rate |

## Corresponding Authors

* viola.krizakova@mat.ethz.ch (V. Krizakova)

* pietro.gambardella@mat.ethz.ch (P. Gambardella)

* kevin.garello@cea.fr (K. Garello)

## Declaration of Competing Interest

The authors declare no competing financial interests.

## Acknowledgments

The authors would like to acknowledge the financial support by the Swiss National Science Foundation (Grant No. 200020_200465), by IMEC's industrial affiliation program on MRAM device, and by the ECSEL joint undertaking program (Grant No 876925 – Project Andante).